\pdfoutput=1
\RequirePackage{ifpdf}
\ifpdf 
\documentclass[pdftex]{sigma}
\else
\documentclass{sigma}
\fi

\begin{document}

\allowdisplaybreaks

\renewcommand{\PaperNumber}{039}

\FirstPageHeading

\ShortArticleName{Drinfeld Doubles for Finite Subgroups of ${\rm SU}(2)$ and ${\rm SU}(3)$ Lie Groups}
\ArticleName{Drinfeld Doubles for Finite Subgroups\\ of $\boldsymbol{{\rm SU}(2)}$ and $\boldsymbol{{\rm SU}(3)}$ Lie Groups}

\Author{Robert COQUEREAUX~$^{\dag\ddag}$ and Jean-Bernard ZUBER~$^\S$}
\AuthorNameForHeading{R.~Coquereaux and J.-B.~Zuber}

\Address{$^\dag$~IMPA \& UMI 2924 CNRS-IMPA,  Jardim Bot\^anico, Rio de Janeiro - RJ, 22460-320, Brazil}

\Address{$^\ddag$~Aix Marseille Universit\'e, CNRS, CPT, UMR 7332, 13288 Marseille, France}
\EmailD{\href{mailto:robert.coquereaux@cpt.univ-mrs.fr}{robert.coquereaux@cpt.univ-mrs.fr}}
\URLaddressD{\url{http://www.cpt.univ-mrs.fr/~coque/}}

\Address{$^\S$~LPTHE, CNRS-UMR 7589 and Universit\'e Pierre et Marie Curie,\\
\hphantom{$^\S$}~4 place Jussieu, 75252, Paris Cedex 5, France}
\EmailD{\href{mailto:zuber@lpthe.jussieu.fr}{zuber@lpthe.jussieu.fr}}

\ArticleDates{Received December 21, 2012, in f\/inal form May 15, 2013; Published online May 22, 2013}

\Abstract{Drinfeld doubles of f\/inite subgroups of SU(2) and SU(3) are investigated in detail.
Their modular data~-- $S$, $T$ and fusion matrices~-- are computed explicitly, and illustrated by means of fusion graphs.
This allows us to reexamine certain identities on these tensor
product or fusion multiplicities
under conjugation of representations  that had been discussed in our recent paper
[{\it J.~Phys.~A: Math. Theor.} {\bf 44} (2011),  295208, 26~pages],
proved to hold for simple and af\/f\/ine Lie algebras, and found to be generally wrong for f\/inite groups. It is shown
here that these identities fail also in general for Drinfeld doubles, indicating that modularity of the fusion
category is not the decisive feature.
Along the way, we collect many data on these
Drinfeld doubles which are interesting for their own sake and maybe
also in a~relation with the theory of orbifolds in conformal f\/ield theory.}


\Classification{81R50; 81T40; 20C99; 18D10}

\vspace{-2mm}

\section{Introduction}

Since their introduction by Drinfeld~\cite{Drinfeld:quasibialgebras,Drinfeld:quasibialgebras1} quantum doubles of Hopf algebras, in
particular of group algebras, have been the subject of some attention, mostly in connection with conformal
f\/ield theory, representation theory, quantum integrability and topological quantum
computation~\cite{CGR:modulardata,DiPaRo:double,Finch-et-al, KacTodorov,KoornEtAl,Ostrik:modulecategoryfordoubles,MThompson:quantumcomputing, Witherspoon}.
Actually the possibility of associating an ``exotic Fourier transform'' (to become the $S$ matrix of
a~quantum double) with any f\/inite group was introduced by Lusztig already in~\cite{Lusztig}, see
also~\cite{Lusztig:exotic}.

The purpose of the present paper is two-fold.
As the Drinfeld doubles of specif\/ic f\/inite groups may be determined quite explicitly, at least when
their order and class number (number of irreps) are not too big, our f\/irst purpose is to invite the
reader to a~tour through a~selection of examples taken from the list of f\/inite subgroups of SU(2) and SU(3).
This is clearly an arbitrary choice, which can be justif\/ied from the interest of those subgroups in the
construction of the so-called orbifolds models in CFT, and also because their modular data and fusion rules
have been determined~\cite{Coq-12}.
Accordingly the reader will f\/ind here a~selected number of data, tables and graphs (in particular fusion
graphs), and we hope that our discussion of the various examples will bring some better understanding in
the study of quantum doubles of f\/inite groups.
More data are available on request and/or on a~dedicated web site\footnote{See
\url{http://www.cpt.univ-mrs.fr/~coque/quantumdoubles/comments.html}.}.

\looseness=-1
Secondly, we want to use these data to explore further an issue that has remained somewhat elusive so far:
in a~recent paper~\cite{RCJBZ:sumrules}, we uncovered identities satisf\/ied by sums of tensor or fusion
multiplicities under complex conjugation of representations.
This is recalled in detail in Section~\ref{section23} below.
These identities were proved for simple Lie algebras of f\/inite or af\/f\/ine type as resulting from
a~case by case analysis, but no conceptually simple interpretation of that result was proposed.
In contrast, these identities were found to fail in a~certain number of f\/inite groups.
It was suggested~\cite{ChSchweigert} that it would be interesting to test these identities on Drinfeld
doubles, as they share with af\/f\/ine Lie algebras the property of being modular tensor categories, in
contradistinction with f\/inite groups.
We shall see below that it turns out that our identities are generally not satisf\/ied by Drinfeld doubles,
which indicates that the modular property is not the decisive factor.

\looseness=-1
The paper is organized as follows.
Section~\ref{section2} recalls a~few facts on orbifolds and Drinfeld doubles; it also displays the expressions of
modular $S$ and $T$ matrix given in the literature~\cite{Coq-12, CGR:modulardata} and reviews the symmetry
properties of $S$, to be used in the following. Section~\ref{subsection24} presents the sum rules of the fusion
coef\/f\/icients and of the $S$ matrix, to be investigated below.
In Section~\ref{section3}, we found it useful to collect a~number of ``well known'' facts on f\/inite groups, that we use
in the sequel.
Finally in Section~\ref{section4}, we come to the explicit case of f\/inite subgroups of~SU(2) and~SU(3).
For each of these two Lie groups, a~short recapitulation of what is known on their f\/inite subgroups is
given, before a~case by case analysis of their Drinfeld doubles.
The discussion is exhaustive for SU(2), whereas for SU(3) we content ourselves with a~detailed discussion
of the ``exceptional'' subgroups and a~few comments on some subgroups of inf\/inite series.
Our results are gathered in Tables~\ref{ssgrSU2} and~\ref{ssgrSU3}.
Finally Appendices collect additional comments on f\/inite groups, explicit modular matrices for the
particular case of the Drinfeld double of the Klein group $\Sigma_{168}$, and \dots\ {a surprise picture}.

\section{Orbifolds, doubles and modular data}\label{section2}

\subsection{Remarks about orbifolds models}

In CFT, a~general orbifold model is specif\/ied by a~pair $({\Gamma},k)$ where ${\Gamma}$ is a~Lie group
and $k$ is a~positive integer, together with a~f\/inite subgroup $G$ of ${\Gamma}$.
When $k=0$, orbifold models (called holomorphic) are actually specif\/ied by a~pair $(G,\omega)$ where $G$
is a~f\/inite group and $\omega$ is a~cocycle belonging to $H^3(G,{\rm U}(1))$.
The group $G$ can be for instance a~subgroup of some given Lie group~${\Gamma}$ (but the latter plays no
role in the construction when $k=0$).
General orbifold models (with $k\neq0$) are discussed in~\cite{KacTodorov}.
These data determine a~f\/inite collection of ``primary f\/ields'', or ``simple objects'', or irreps, for
short.
It also determines a~fusion ring linearly spanned by the irreps.
Those are the so-called chiral data on which one builds a~CFT, here an orbifold theory.
 In BCFT, one has further to specify the kind of ``boundary'' one considers.
A specif\/ic boundary type determines a~module (also called nimrep) over the fusion ring~\cite{BPPZ, Ca}.

Using the vocabulary of category theory, one may say that an orbifold model def\/ines a~fusion category
${\mathcal A}$ whose simple objects are the primary f\/ields $\sigma_n$, labelled by $n$ running from $1$ to~$r_A$.
The fusion ring is generated linearly by the primary f\/ields $\sigma_n$.
The ring structure $\sigma_m \sigma_n=\sum\limits_p N_{mn}^{\phantom{m}p}\sigma_p$ is specif\/ied
by non-negative integers $N_{mn}^{\phantom{m}p}$.
The category is modular: we have~$S$ and~$T$ matrices, of dimensions $r_A\times r_A$, representing the
generators of the group ${\rm SL}(2,\mathbb{Z})$; they obey $S^2=(ST)^3=C$, $C^2=\ensuremath{\,\,\mathrm{l\!\!\!1}}$.
The matrix $C$ is called the conjugation matrix.
{$S$ and $T$ are unitary, $S$ is symmetric.} The fusion coef\/f\/icients $N_{mn}^{\phantom{m}p}$ are given by the
Verlinde formula~\cite{Verlinde}{\samepage
\begin{gather}
\label{Verl}
N_{mn}^{\phantom{m}p}=\sum_{\ell}\frac{S_{m\ell}S_{n\ell}S_{p\ell}^{*}}{S_{1\ell}}.
\end{gather}
In the case $k=0$ (level $0$) the Lie group ${\Gamma}$ plays no role, and we set ${\mathcal A}={\mathcal A}(G,\omega)$.}

A BCFT def\/ines (or is def\/ined by) a~module-category ${\mathcal E}$ over ${\mathcal A}$.
Its simple objects (called boundary states, in BCFT), labelled by $a$ from $1$ to $r_E$ are denoted
$\tau_a$.
They generate an abelian group which is a~module over the ring of ${\mathcal A}$.
Explicitly, $\sigma_m\,\tau_a=\sum\limits_b F_{m,a}^{\phantom{m,}b}\tau_b$.
The constants $F_{m,a}^{\phantom{m,}b}$
are also non-negative integers (whence the acronym nimreps).
A BCFT (a choice of ${\mathcal E}$) is associated with a~symmetric matrix $Z$, of dimensions $r_A\times
r_A$, also with non-negative integer coef\/f\/icients, that commutes with $S$ and $T$.
For this reason $Z$ is called the modular invariant matrix.
It is normalized by the condition $Z_{1,1}=1$.
For the particular choice ${\mathcal E}={\mathcal A}$, the boundary states coincide with the primary
f\/ields, $r_E=r_A$, and the modular invariant matrix $Z=\ensuremath{\,\,\mathrm{l\!\!\!1}}$ is the unit matrix.
The construction of $Z$ from the BCFT/module category ${\mathcal E}$ or vice versa remains in practice
a~matter of art \dots.

In general, from any f\/inite group $G$, one can build ${\mathcal A}(G)={\mathcal A}(G,0)$ by the so-called
Drinfeld double construction \cite{Drinfeld:quasibialgebras,Drinfeld:quasibialgebras1}: it is the representation category of a~Hopf
algebra {$D(G)$} called the Drinfeld double of $G$, or the (untwisted) quantum double of $G$.
More generally, from any f\/inite group $G$, together with a~cocycle\footnote{In $H_3(G,\mathbb{Z})\cong
H^3(G,\mathbb{C}^\times)\cong H^3(G,{\rm U}(1))$.} $\omega$, one can build a~fusion category ${\mathcal
A}(G,\omega)$ by a~method called the twisted Drinfeld double construction.
The genuine Hopf algebra $D(G)$ is replaced by a~quasi-Hopf algebra $D_\omega(G)$.
The latter is a~quasi-bialgebra, not a~bialgebra, because the coproduct is not co-associative.

\begin{remark}
One may often use various {methods} to build the same category ${\mathcal A}$, up to equivalence.
The Hopf algebra $D(G)$ and the twisted Hopf algebras $D_\omega(G)$ have been used in~\cite{DiPaRo:double}
to build~${\mathcal A}(G)$ and~${\mathcal A}(G,\omega)$, but other constructions should be possible.
\end{remark}

\looseness=-1
According to~\cite{Ostrik:modulecategoryfordoubles} the indecomposable\footnote{Not equivalent to a~direct
sum.} nimreps ${\mathcal E}$ of ${\mathcal A}(G,\omega)$ or in other words the indecomposable
module-categories ${\mathcal E}$ over ${\mathcal A}(G,\omega)$ are parametrized by the conjugacy classes of
pairs $(K,\psi)$ where $K\subset G\times G$ is a~subgroup, $\psi$ a~cohomology class in
$H^2(K,\mathbb{C}^\times)$, and $K$ is such that the natural
extension\footnote{$\tilde\omega=p_1^\star\omega-p_2^\star\omega$, where $p_i$ are projections $G\times
G\to G$: $(g_1,g_2)\mapsto g_i$.} $\tilde\omega$ of the cohomology class $\omega$ to $H^3(G\times
G,\mathbb{C}^\times)$ is trivial on $K$.
Such subgroups $K$ of $G\times G$ are called admissible for $(G,\omega)$.
This latter freedom, that usually (not always) changes the modular invariant partition function but not the
modular data, was called ``discrete torsion'' in~\cite{Vafa:discretetorsion}, and in~\cite{CGR:modulardata}.
It is clear that any subgroup $K$ of $G\times G$ is admissible for $\omega=0$.

In what follows we shall only consider holomorphic orbifolds, and moreover often assume that the cocycle
$\omega$ is trivial (in other words we shall consider ``untwisted holomorphic orbifolds'').
For this reason, we shall write ``Drinfeld double'' instead of ``quantum double'' in the following.
Moreover, we shall not discuss boundary states, BCFT, nimreps, and the like~$\ldots$. Nevertheless, we
believe that it was not useless to remind the reader of the above facts concerning module-categories
associated with orbifold models, in order to better understand ``where we stand''!

So, what matters for us in this paper is mostly the (modular) fusion category ${\mathcal A}(G)$ associated
with the choice of a~f\/inite group $G$.
It will actually be enough to know how matrices~$S$ and~$T$ are constructed from some f\/inite group data
(see formulae~\eqref{ST-formulas} below).
The multiplicative structure (the fusion ring coef\/f\/icients $N_{mn}^{\phantom{m}p}$) can be obtained from $S$ via Verlinde equations~\cite{Verlinde}.
In particular the Drinfeld double construction (twisted or not), {which may be used to obtain the general
formulae in Section~\ref{formulaForS} will not be explicitly used in the sequel}.
In what follows we shall only consider fusion rings obtained from Drinfeld doubles of f\/inite groups, and
we therefore drop the general notation ${\mathcal A}(G)$ and write $D(G)$ instead.

\subsection{General properties of Drinfeld doubles}
\label{formulaForS}
\begin{itemize}\itemsep=0pt

\item We shall call ``rank'' $r$ the total number of irreps of $D(G)$.
As the irreps of $D(G)$ are labelled by pairs $({[c]},\sigma_{c})$, where ${[c]}$ is a~conjugacy class of
the group $G$ and $\sigma_{c}$ an irrep of the centralizer in $G$ of (any representative ${c}$ of)
${[c]}$,\footnote{Two elements of the same conjugacy class have isomorphic centralizers.} we group together in ``blocks'' those associated with the same conjugacy class and centralizer.
For each example we shall list the number $N_c$ of elements (i.e.\ of irreps $\sigma_{c}$) in each block~$c$.
Their sum is thus equal to the rank $r$.
We call ``classical'' those irreps of the Drinfeld double that correspond to the f\/irst block called
``classical block'', associated with the trivial conjugation class (the centralizer of the identity being
$G$ itself, these irreps can be identif\/ied with the irreps of the group~$G$).
Their number is equal to the number of conjugacy classes of $G$, that we call the class number.

\item Quantum dimensions of irreps, for fusion models built from Lie groups at f\/inite levels (WZW
theories), are usually not integers, but quantum dimensions of irreps of doubles of f\/inite groups are
always integers.
When those irreps are classical, their quantum dimensions coincide with the dimensions of the corresponding
irreps of the group.

\item If $\chi=(c,\sigma_c)$ is an irrep of $D(G)$, its quantum dimension is $\mu(\chi)$, and the global
dimension of $D(G)$ is def\/ined as $\vert D(G)\vert=\sum \mu(\chi)^2$.
In the case of Drinfeld doubles, where the cocycle is trivial, we have $\vert D(G)\vert=|G|^2$, where $|G|$
is the order of $G$.

\item For each of the examples that we consider later, we shall also give the integer $d_{\mathcal
B}=\sum\limits_m(\sum\limits_{n,p}N_{mn}^{\phantom{m}p})^2$ whose interpretation as the dimension of a~weak Hopf algebra ${\mathcal B}$ (or double triangle
algebra~\cite{Ocneanu:Fields}) will not be discussed in this paper,  see also~\cite{CoquereauxIsasiSchieber:TQFT, Hayashi, Ostrik,
PetkovaZuber:cells}.

\item In writing the $S$, $T$ and fusion matrices, we sort the irreps as follows.
First of all we sort the conjugacy classes according to the increasing order $p$ (in the sense $g^p=1$) of
its representatives.
For instance the conjugacy class of the identity (for which $p=1$) always appears f\/irst.
Whenever two classes have the same $p$, their relative order is arbitrarily chosen.
Finally, for a~given conjugacy class, the irreps of the associated centralizer are ordered (not totally)
according to their increasing classical dimension.

\item Formulae for $S$ and $T$:

We copy from~\cite{CGR:modulardata} the following expressions for the untwisted case.
More general expressions that are valid for any twist can be found in the same reference, where they are
used to explicitly determine the corresponding $S$ and $T$ matrices in several cases, in particular for the
odd dihedral groups (any twist).
As recalled above, there is a~one to one correspondence between irreps of $D(G)$ and pairs
$([c],\sigma_c)$, where $[c]$ is a~conjugacy class of $G$, and $\sigma_c$ denotes an irrep of the
centralizer $C_G(c)$ of (any representative $c$ of) class $[c]$ in $G$.
Then
\begin{gather}
\nonumber
S_{([c],\sigma_c)([d],\sigma_d)}=\frac{1}{|C_G(c)||C_G(d)|}\sum_{\scriptstyle{g\in G}
\atop\scriptstyle{c g d g^{-1}=g d g^{-1}c}}\chi_{\sigma_c}\big(g d g^{-1}\big)^{*}\chi_{\sigma_d}\big(g^{-1}cg\big)^{*}
\\
\label{ST-formulas}
\phantom{S_{([c],\sigma_c)([d],\sigma_d)}}
=\frac{1}{|G|}\sum_{g\in[c],h\in[d]\cap C_G(c)}\chi_{\sigma_c}\big(x h x^{-1}\big)^{*}\chi_{\sigma_d}\big(ygy^{-1}\big)^{*},
\\
T_{([c],\sigma_c)([d],\sigma_d)}=\delta_{cd}\delta_{\sigma_c\sigma_d}\frac{\chi_{\sigma_c}(c)}
{\chi_{\sigma_c}(e)},
\nonumber
\end{gather}
where $x$ and $y$ are arbitrary solutions of $g=x^{-1}cx$ and $h=y^{-1}dy$.

In practice, it is more convenient to use a~variant of~\eqref{ST-formulas}~\cite{Coq-12}.
Let ${\mathcal T}_c=\{c_i\}$ (resp.\ ${\mathcal T}_d=\{d_j\}$) be a~system of coset representatives for the
{left} classes of~$G/C_G(c)$ (resp.\ a system of coset representatives for the left classes of~$G/C_G(d)$),
then
\begin{gather}
\label{S-alt}
S_{([c],\sigma_c)([d],\sigma_d)}=\frac{1}{\vert G\vert} \sum_{c_i,d_j\atop g_{ij}=c_i d_j^{-1}} \!\!\!\!\!\!\! {\vphantom{\sum}}' \; \chi_{{\sigma_c}}\big(g_{ij}d g_{ij}^{-1}\big)^{*}\chi_{{\sigma_d}}\big(g_{ij}^{-1}cg_{ij}\big)^{*},
\end{gather}
where the primed sum runs over pairs of $c_i\in{\mathcal T}_c$, $d_j\in{\mathcal T}_d$ that obey
$[b_j^{-1}b b_j,a_i^{-1}a~a_i]=1$; here $[{}\;,\,{}]$ is the commutator def\/ined as
$[a,b]=a^{-1}b^{-1}ab$.
This reformulation of~\eqref{ST-formulas}, also used implicitly in~\cite{MThompson:quantumcomputing}, is
handy because sets of coset representatives are provided by GAP~\cite{GAP}.
\end{itemize}

\subsection[Symmetries of the $S$ matrix]{Symmetries of the $\boldsymbol{S}$ matrix}\label{section23}
\begin{itemize}\itemsep=0pt
\item The most conspicuous property of the $S$-matrix in~\eqref{ST-formulas} or~\eqref{S-alt} is its
symmetry:
\begin{gather*}
{S_{([c],\sigma_c)([d],\sigma_d)}=S_{([d],\sigma_d)([c],\sigma_c)}.}
\end{gather*}
Compare with the case of an ordinary f\/inite group $G$, for which the tensor product multiplicities are
given by
\begin{gather}\label{tensor-group}
N_{rs}^{\phantom{r}t}=\sum_c\frac{\hat\chi_c^{\phantom{c}r}\hat\chi_c^{\phantom{c}s}\hat\chi_c^{\phantom{c}t*}}{\hat\chi_c^{\phantom{c}1}},
\end{gather}
where $\hat\chi_c^{\phantom{c}r}=\sqrt{\frac{|c|}{|G|}}\chi_c^{\phantom{c}r}$ is the normalized character of irrep $r$ in
class $c$, an expression which looks like Verlinde formula~\eqref{Verl}.
In that case, however, there is no reason that the diagonalizing matrix $\hat\chi$ of multiplicities be
symmetric, and it is generically not.
In contrast in a~Drinfeld double, that matrix, called now $S$, is symmetric.
In other words, there is not only an equal number of classes and irreps in a~double, there is also
a~canonical bijection between them.
The origin of that symmetry may be found in a~CFT interpretation~\cite{DiPaRo:double}, or alternatively,
may be derived directly~\cite{KoornEtAl}.
\item The $S$-matrix has other properties that are basic for our purpose:
\begin{itemize}\itemsep=0pt
\item it is unitary, $S.S^\dagger=I$; \item its square $S^2=C$ satisf\/ies $C.C=I$, i.e.~$S^4=I$.
As recalled above, this is, with $(S.T)^3=C$, one of the basic relations satisf\/ied by generators of the
modular group.
Since $S^{*}=S^\dagger=S^3=S.C=C.S$, the matrix $C$ is the conjugation matrix,
$C_{ij}=\delta_{i\bar\jmath}$.
\end{itemize}\itemsep=0pt
\item As just mentioned, under complex conjugation, $S$ transforms as
\begin{gather*}
S_{ij}^{*}=S_{\bar\imath j}=S_{i\bar\jmath},
\end{gather*}
where $\bar\imath$ refers to the complex conjugate irrep of $i$; in the case of the double, where $i$
stands for $([c],\sigma_c)$, $\bar\imath$ stands for
$\overline{([c],\sigma_c)}=([c^{-1}],\overline{\sigma_c})$.
This follows from the formulae in~\eqref{ST-formulas} and~\cite{CGR:modulardata}.
By Verlinde formula this implies that
\begin{gather}
\label{conj-fusion}
N_{\bar\imath}=N_i^T.
\end{gather}
Thus, for tensor product (fusion), complex conjugation amounts to transposition, a~pro\-perty also enjoyed
by~\eqref{tensor-group}.
Moreover, $N_{\bar\imath}=C.N_i.C$.
\item Other symmetries of the $S$ matrix of the double are associated with the existence of units in the
fusion ring.
An invertible element in a~ring is usually called a~unit.
A fusion ring is a~$\mathbb{Z}_+$ ring,  i.e.,  it comes with a~$\mathbb{Z}_+$ basis (the irreps),
and in such a~context, one calls units those irreps that are invertible (in the context of CFT, units are
generally called {\it simple currents}).
Therefore if $u$ is a~unit, hence an irrep such that $N_u$ is invertible, necessarily $N_u^{-1}=N_{\bar
u}=N_u^T$, $N_u$ is an orthogonal matrix and $\det N_u=\pm1$.
In view of~\eqref{conj-fusion}, $N_u$ is an orthogonal integer-valued matrix, hence a~permutation matrix,
$(N_u)_i^{\phantom{i}j}=\delta_{J_u(i),j}$, where $J_u$ is a~permutation.
\item In the following we denote
\begin{gather}
\label{def-phi}
\phi_i(\ell)=\frac{S_{i\ell}}{S_{1\ell}}
\end{gather}
the eigenvalues of the $N_i$ fusion matrix.
Note that for a~unit $u$, $\phi_u(\ell)$ is a~root of unity.
Moreover, as $\phi_u(1)$ is also a~quantum dimension, hence a~positive number, this must necessarily be
equal to~1.
\item The existence of units entails the existence of symmetries of the {{\it fusion graphs}, also called
{\it representation graphs} in the literature, namely the graphs\footnote{All the graphs given in this
paper, as well as many calculations involving fusion matrices, have been obtained with the symbolic package
Mathematica~\cite{Mathematica}, interfaced with GAP~\cite{GAP}.} whose adjacency matrices are the fusion
matrices.} Each permutation $J_u$, for $u$ a~unit, acts on irreps in such a~way that
\begin{gather*}
\forall\, i
\quad
N_i=N_u N_{\bar u}N_i=N_u N_i N_{\bar u}\ \Rightarrow \ N_{ij}^{\phantom{i}k}=N_{iJ_u(j)}^{\phantom{i}J_u(k)},
\end{gather*}
hence may be regarded as an automorphism of the fusion rules and a~symmetry of the fusion graphs: on the
fusion graph of any $N_i$, there is an edge from $j$ to $k$ if\/f there is one from $J_u(j)$ to $J_u(k)$.
A particular case of such automorphisms is provided by the automorphisms of weight systems in af\/f\/ine
algebras used in~\cite{RCJBZ:sumrules}.
\item As all irreps of the double, units are labelled by pairs $([c],\psi)$, but here the class $[c]$ is
the center $Z(G)$ of $G$, its centralizer is $G$ itself, and $\psi$ is a~1-dimensional irrep of $G$.
Indeed, for Drinfeld doubles, the quantum dimension $S_{([c],\sigma_c)([e],1)}/S_{([e],1)([e],1)}$ of an
irrep $j=([c],\sigma_c)$ is equal to $\vert[c]\vert\times \dim(\sigma_c)$, but for a~unit
$([c],\psi=\sigma_c)$, the quantum dimension is equal to~$1$ (see above after~\eqref{def-phi}), and $c$ is
central ($\vert[c]\vert=1$), so $\psi$ is of degree $1$.
The set of the latter is given by the `abelianization' $G/G'$ of $G$, with $G'$ the commutator subgroup of~$G$.
Thus the group of units is isomorphic to $Z(G)\times G/G'$.
\end{itemize}

\subsection[Sum rules for the $S$ matrix]{Sum rules for the $\boldsymbol{S}$ matrix}\label{subsection24}

Let $N_{ij}^{\phantom{i}k}$ stand for the multiplicity of irrep $k$ in $i\otimes j$ and let $\bar\imath$ refer to the complex
conjugate irrep of $i$.
According to~\cite{RCJBZ:sumrules}, both in the case of semi-simple Lie groups and in the case of fusion categories
def\/ined by a~pair ${(\Gamma,k)}$ (WZW models), we have
\begin{gather}
\label{sumrule}
\forall\,  i,j
\qquad
{\sum_k N_{ij}^{\phantom{i}k}{=}\sum_k N_{\bar\imath j}^{\phantom{i}k}.
}
\end{gather}
or equivalently
\begin{gather}\label{sumrule'}
\forall\,  j,k
\qquad
\sum_i N_{ij}^{\phantom{i}k}{=}\sum_i N_{i j}^{\phantom{i}\bar k}.\tag{$7'$}
\end{gather}

In the case of WZW models, where the category is modular, we have shown the above property to be equivalent
to the following: if an irrep $j$ is of complex type, then
\begin{gather}
\label{sumruleS}
 \Sigma_j:=\sum_i S_{ij}=0,
\end{gather}
and we shall say below that the irrep $j$ has a~vanishing $\Sigma$.
Actually we have shown in~\cite{RCJBZ:sumrules} that the last property also holds when $j$ is of
quaternionic type.

Def\/ining the charge conjugation matrix $C=S^2$ and the path matrix $X=\sum\limits_i N_i$, it is
a~standard fact that $C.X.C=X$.
Property \eqref{sumrule'} reads instead
\begin{gather}
\label{XeqXC}
X=X.C=C.X.
\end{gather}

The f\/irst natural question is to ask whether property (9) holds for f\/inite groups.
As noticed in~\cite{RCJBZ:sumrules}, the answer is {in general} negative (although it holds {in many
cases}).
To probe equation~\eqref{sumrule}, we have to look at groups possessing complex representations.
In the case of ${\rm SU}(2)$ subgroups, equation~\eqref{sumrule} holds, and this was easy to check since
only the cyclic and binary tetrahedral subgroups have complex representations.
It was {then} natural to look at subgroups of ${\rm SU}(3)$ and we found that {\eqref{sumrule}} holds true
for most subgroups of ${\rm SU}(3)$ but fails for some subgroups like $F=\Sigma_{72\times3}$ or
$L=\Sigma_{360\times3}$.
The second property~\eqref{sumruleS} does not make sense for a~f\/inite group since there is no invertible
$S$ matrix, and Verlinde formula cannot be used.

The next natural question\footnote{We thank Ch.~Schweigert for raising that issue.} is to ask if the above properties~\eqref{sumrule}, \eqref{sumruleS}
hold for Drinfeld doubles of f\/inite groups.
As we shall see in a~forthcoming section, the answer is again negative.

Let us now prove now the following

\begin{proposition}\label{proposition1}
For a~Drinfeld double:
$($equation~\eqref{sumrule}$)$ $\Leftrightarrow$ $\forall\, j\ne\bar j$, $\Sigma_j=0$.
\end{proposition}

Our proof follows closely the steps of the proof of a~similar statement in~\cite{RCJBZ:sumrules} in the
case of f\/inite dimensional or af\/f\/ine Lie algebras, although here neither property is necessarily valid.
\begin{itemize}\itemsep=0pt
\item~\eqref{sumruleS} $\Rightarrow$~\eqref{sumrule}.
Suppose that only self-conjugate irreps have a~non vanishing $\Sigma$ and use~\eqref{Verl} to write
\begin{gather*}
\sum_k N_{ij}^{\phantom{i}k}=\sum_\ell\frac{S_{i\ell}S_{j\ell}\sum_k S^{*}_{k\ell}}{S_{1\ell}}=\sum_{\ell=\bar\ell}
\frac{S_{i\ell}S_{j\ell}\sum_k S^{*}_{k\ell}}{S_{1\ell}}
=\sum_k\sum_{\ell=\bar\ell}
\frac{S_{\bar\imath\ell}S_{j\ell}S^{*}_{k\ell}}{S_{1\ell}}=\sum_k N_{\bar\imath j}^{\phantom{\bar\imath}k}.
\end{gather*}
\item \eqref{sumrule} $\Rightarrow$~\eqref{sumruleS}.
Suppose that $\sum\limits_i N_{ij}^{\phantom{i}k}=\sum\limits_i N_{i j}^{\phantom{i}\bar k}$ for all $i$, $j$, $k$.
Use again~\eqref{Verl} and~\eqref{sumrule} to write
\begin{gather*}
\left(\sum_i S_{i\ell}\right)S_{j\ell}=\sum_k\sum_i N_{ij}^{\phantom{i}k}S_{k\ell}S_{1\ell}=\sum_k\sum_i N_{ij}
^{\phantom{i}\bar k}S_{k\ell}S_{1\ell}
\\
\phantom{\left(\sum_i S_{i\ell}\right)S_{j\ell}}
=\sum_k\sum_i\! N_{ij}^{\phantom{i}k}S_{\bar k\ell}S_{1\ell}=\sum_k\sum_i\!  N_{ij}
^{\phantom{i}k}S_{k\bar\ell}S_{1\bar\ell}=\sum_i\!  S_{i\bar\ell}S_{j\bar\ell}= \left(\sum_i S_{i\ell}\right) \! S_{j\bar\ell},
\end{gather*}
from which we conclude that if $\sum\limits_i S_{i\ell}\ne0$, then $S_{j\ell}=S_{j\bar\ell}$, which cannot
hold for all~$j$ unless $\ell=\bar\ell$ (remember that $S_{j\ell}/S_{1j}$ and $S_{j\bar\ell}/S_{1j}$ are
the eigenvalues of two fusion matrices~$N_\ell$ and $N_{\bar\ell}$ which are dif\/ferent if
$\ell\ne\bar\ell$).
Thus, assuming~\eqref{sumrule} (which is not always granted), if $\ell\ne\bar\ell$, $\sum\limits_i
S_{i\ell}=0$. 
\end{itemize}

\begin{proposition}\label{proposition2}
In any modular tensor category, the complex conjugation is such that
properties~\eqref{sumrule} and~\eqref{sumruleS} are simultaneously true or wrong.
\end{proposition}

\begin{remark}
As proved in~\cite{RCJBZ:sumrules}, property~\eqref{sumrule} hold in the case of Lie groups, and
for af\/f\/ine Lie algebras at level $k$ (WZW models), both properties~\eqref{sumrule} and~\eqref{sumruleS}
hold.
In the case of Drinfeld doubles of f\/inite groups, it is not always so.
\end{remark}

 Whenever the fusion/tensor ring has units, we may state the following

\begin{proposition}\label{proposition3}
Consider an irrep $j$ such that there exists a~unit $u$ with
\begin{gather*}
\phi_u(j)=\frac{S_{uj}}{S_{1j}}\ne1.
\end{gather*}
Then~\eqref{sumruleS} holds true: $\sum\limits_i S_{ij}=0$.
\end{proposition}

We write simply, using the fact that $N_u=J_u$ is a~permutation
\begin{gather*}
\nonumber
\sum_i S_{ij}=\sum_i S_{J_u (i)j}=\phi_u(j)\sum_i S_{ij}
\end{gather*}
and $\phi_u(j)\ne1\Rightarrow\sum\limits_i S_{ij}=0$. 

One f\/inds, however, cases of complex irreps $j$ for which all units $u$ give $\phi_u(j)=1$ and
Proposition~\ref{proposition3} cannot be used.
In the examples (see below), we shall encounter the two possibilities:
\begin{itemize}\itemsep=0pt
\item Complex irreps (with all $\phi_u(j)=1$) such that $\Sigma_j\ne0$, hence counter-examples to
pro\-per\-ty~\eqref{sumruleS}.
\item Vanishing $\Sigma_{j}$ (cf.~\eqref{sumruleS}) for complex, quaternionic and even real irreps $j$ for
which all \mbox{$\phi_u(j)=1$}.
We call such cases ``accidental cancellations'', by lack of a~better understanding.
\end{itemize}

\section{Finite group considerations}\label{section3}

\subsection{About representations, faithfulness, and embeddings}\label{subsection31}

Before embarking into the study of Drinfeld doubles for f\/inite subgroups of ${\rm SU}(2)$ and ${\rm SU}(3)$,
we need to introduce some terminology and remind the reader of a~few properties that belong to the
folklore of f\/inite group theory but that we shall need in the sequel.

Any faithful unitary $n$-dimensional (linear) representation of a~f\/inite group $G$ on a~complex vector
space def\/ines an embedding of $G$ into ${\rm U}(n)$.
An extra condition (the determinant of the representative group elements should be $1$) is required in
order for $G$ to appear as a~subgroup of ${\rm SU}(n)$.
Let us assume, from now on, that $G$ is a~subgroup of ${\rm SU}(n)$.
When the chosen $n$-dimensional representation def\/ining the embedding is irreducible, $G$ itself is
called irreducible with respect to ${\rm SU}(n)$.
When $n>2$, we call {\it embedding representations} with respect to ${\rm SU}(n)$, those irreps that are
$n$-dimensional, irreducible and faithful.
The type of a~representation can be real, complex, or quaternionic.
As the fundamental (and natural) representation of ${\rm SU}(2)$ is quaternionic, we adopt in that case
a~slighly more restrictive def\/inition.
For a~f\/inite group~$G$, isomorphic with an irreducible subgroup of ${\rm SU}(2)$, we call {\it embedding
representations} with respect to ${\rm SU}(2)$, those irreps that are $2$-dimensional, irreducible,
faithful and of quaternionic type.
More details can be found in Appendix~\ref{appendixA}.

At times, we shall need the following notion.
A f\/inite subgroup $G$ of a~Lie group is called Lie primitive (we shall just write ``primitive'') if it is
not included in a~proper closed Lie subgroup.
Although irreducible with respect to their embedding into ${\rm SU}(3)$, some of the subgroups that we
shall consider later are primitive, others are not.
More details can be found in Appendix~\ref{appendixB}.

The fundamental representation of dimension $3$ of ${\rm SU}(3)$, or its conjugate, is usually called the
natural (or def\/ining) representation, and it is faithful.
In this paper we shall mostly, but not always, consider subgroups that are both irreducible and primitive,
and the given notion of embedding representation is appropriate (it may be non-unique, see below).
However, in some cases, the previous notion of embedding representation should be amended.
This is in particular so for the cyclic subgroups of ${\rm SU}(2)$ where no irreducible, faithful,
2-dimensional, and of quaternionic type exists.
Notice that for some ${\rm SU}(3)$ subgroups, there are cases where the embedding representation~-- as
def\/ined previously~-- is not of complex type, for the reason that no such representation exists: see
below the examples of of $\Delta(3\times2^2)$, $\Delta(6\times2^2)$, and $\Sigma(60)$, the tetrahedral,
octahedral, and icosahedral subgroups of ${\rm SO}(3)\subset{\rm SU}(3)$, which have no {\it complex}
3-dimensional irrep.
They are not primitive subgroups of ${\rm SU}(3)$.

In the present paper we are not so much interested in fusion graphs associated with f\/inite groups $G$,
rather we are interested in their Drinfeld doubles $D(G)$.
There is one fusion graph for each irrep of $D(G)$ and it is of course out of question to draw all of them
in this paper.
For this reason, we are facing the problem of which representation to select.
As recalled above, irreps of~$D(G)$ are labeled by pairs (a conjugacy class of~$G$ and an irrep of the
corresponding centralizer).
One special conjugacy class is the class of the neutral element $\{e\}$ of $G$, which has only one element,
its centralizer being $G$ itself.
As we are mostly interested in irreducible embeddings that def\/ine f\/inite subgroups~$G$ of~${\rm SU}(n)$
for $n=2$ or~$3$, the selected representation of~$D(G)$ will {(with a~few exceptions, see later)} be of the
type $(\{e\},\rho)$ with $\rho$ chosen among the irreducible faithful representations of~$G$ of the
appropriate dimension $n$.
We shall call ``embedding irrep'' of the Drinfeld double of $G$, any pair $(\{e\},\rho)$ where $\rho$ is an
embedding representation for $G$, with the above meaning.

\subsection{About faithfulness, and connectedness of fusion graphs}

Faithfulness of the selected embedding representation (of~$G$) can be associated with several concepts and
observations related to connectedness properties of the associated fusion graph of~$G$ or of~$D(G)$: the
former is connected whereas the latter appears to have a~number of connected components equal to the class
number of~$G$.

Fundamental representations of a~simple complex Lie group or of a~real compact Lie group can be def\/ined
as those irreps whose highest weight is a~fundamental weight.
These irreps generate by fusion (tensor product) all irreps.
Still for Lie groups, and more generally, we may ask for which irreducible representation $\rho$, if any,
fundamental or not, can we obtain each irreducible representation as a~subrepresentation of a~tensor power
$\rho^{\otimes k}$, for some~$k$.
This is a~classical problem and the answer is that $\rho$ should be faithful~\cite{Huang}; in the same
reference it is shown that, except when $G={\rm Spin}(4n)$, there exist faithful irreducible
representations for all the simple compact Lie groups.

In the theory of f\/inite groups, there is no such notion as being fundamental, for an irreducible
representation.
However, one can still ask the same question as above, and the result turns out to be the same (Burnside,
as cited in~\cite{Curtis-Reiner}): if $G$ is a~f\/inite group and $\rho$ is a~faithful representation of~$G$, then every irreducible representation of $G$ is contained in some tensor power of~$\rho$.
In other words, the fusion graph associated with a~faithful representation of a~f\/inite group is
connected, since taking tensor powers of this representation amounts to following paths on its fusion
graph, and all the irreps appear as vertices.

Let $H$ be a~subgroup of the f\/inite group $G$ and let $\rho$ be a~faithful representation of~$G$.
Then~$\rho_H$, the restriction of $\rho$ to~$H$, may not be irreducible, even if $\rho$ is, but it is
clearly faithful: its kernel, a~subgroup of~$H$, is of course trivial since the kernel of $\rho$ was
already trivial in the f\/irst place.
Therefore every irreducible representation of~$H$ is contained in some tensor power of~$\rho_H$.
Writing $\rho_H a =\sum\limits_b F_{\rho,a}^{\phantom{\rho,}b} b$, where $a$, $b$, $\ldots$ are irreps of $H$, def\/ines a~matrix $F_\rho$ which is the adjacency matrix of a~graph.
This (fusion) graph is connected, for the same reason as before.
Notice that $\rho_H$ itself may not appear among its vertices since it may be non irreducible.

As mentioned previously every representation $\rho$ of $G$ determines a~representation $(e,\rho)$ of~$D(G)$.
The representation rings for the group $G$ and for the algebra~$D(G)$ are of course dif\/ferent, the fusion
coef\/f\/icients of the former being obtained from its character table, those of the latter from the
modular $S$-matrix and the Verlinde formula, but the former can be consi\-de\-red as a~subring of the latter.
Since irreps of the double fall naturally into blocks indexed by conjugacy classes, we expect that the
fusion graph of an embedding irrep of~$D(G)$ will have several connected components, one for each conjugacy
class,  i.e.,  a~number of components equal to the number of classes of~$G$,  i.e.,  to the
class number.
This graph property is actually expected for all the irreps of $D(G)$ stemming from $(z,\rho)$, with
$z\in Z(G)$ and a~faithful irreducible representation $\rho$ of $G$.
Indeed, the usual character table of $G$ can be read from the $S$ matrix in the following way: extract from
$S$ the submatrix made of its f\/irst $\ell$ rows (the ``classical irreps''~$r$), in the latter keep only
the f\/irst column of each of the $\ell$ blocks (corresponding to dif\/ferent classes~$c$ of~$G$) and
f\/inally multiply these columns by ${|G|/|c|}$, resp.\ $\sqrt{{|G|/|c|}}$; this yields the matrix
$\chi^{\phantom{c}r}_c$, resp.\ $\hat\chi^{\phantom{c}r}_c$ def\/ined in~\eqref{tensor-group}.
A similar construction applies to the character tables pertaining to the dif\/ferent centralizers of
conjugacy classes, which may also be extracted from the $S$-matrix~-- the latter is much more than a~simple
book-keeping device for the character tables of the dif\/ferent centralizers of conjugacy classes since it
couples these dif\/ferent blocks in a~non-trivial way.
On the basis of all examples that we have been considering, and in view of the above discussion, we
conjectured

{\it The fusion graph of an embedding irrep of $D(G)$ has $\ell$ connected components, with $\ell$, the
class number, equal to the number of irreps or of conjugacy classes of $G$.}

A formal proof of this property, that we shall not give\footnote{Note added: As noticed by an anonymous
referee to whom we are deeply indebted, such a~proof actually follows from simple considerations making use
of results of~\cite{Cib, DGNO}, in particular of the formula
$(e,\rho)\otimes(a,\delta)=(a,\rho\downarrow^G_{C_G(a)}\otimes \delta)$ where $\rho$ is an irrep of $G$ and
$\delta$ an irrep of the centralizer $C_G(a)$ of $a\in G$.}, can exploit, in the language of fusion
categories, the relation between the representation rings of $G$ and $D(G)$, in particular a~generalization
(see for instance~\cite{Witherspoon}) of the mechanism of induction and restriction.
Notice that an embedding fusion graph, for the group~$G$ (the fusion graph of an embedding representation)
can be obtained by selecting the connected component of~$(e,1)$ in the graph of the corresponding embedding
graphs of its double.
Since it describes a~faithful representation, the number of vertices of this connected component is also
equal to the class number of~$G$.

\subsection{Additional remarks}

In general, a~f\/inite group $G$ may have more than one irreducible
faithful representation of given dimension $n$ (up to conjugation, of course); for instance
a~representation of complex type will always appear together with its complex conjugate in the table of
characters, but there are groups that possess several conjugated pairs of inequivalent faithful irreps of
complex type, with the same dimension, and they may also possess self-dual (i.e.,  real or
quaternionic) faithful irreps, also of the same dimension, see the example of
$\Sigma_{168}\times\mathbb{Z}_3$ below.
There are even f\/inite groups that have more than one pair of faithful irreducible representations of
complex type, with dif\/ferent dimensions, for instance the ${\rm SU}(3)$ subgroup
$\Sigma_{168}\times\mathbb{Z}_3$ possesses faithful irreps of complex type in dimensions $3$, $6$, $7$, $8$.
With the exception of cyclic groups, all the f\/inite subgroups of ${\rm SU}(2)$ have at least one
$2$-dimensional irreducible faithful representation of quaternionic type.
\begin{figure}[htp]\centering
\includegraphics[scale=0.45]{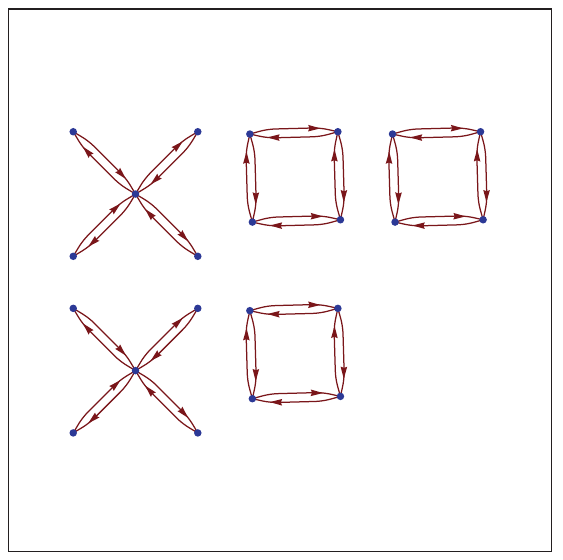} \hspace{4.0cm} \includegraphics[scale=0.5]{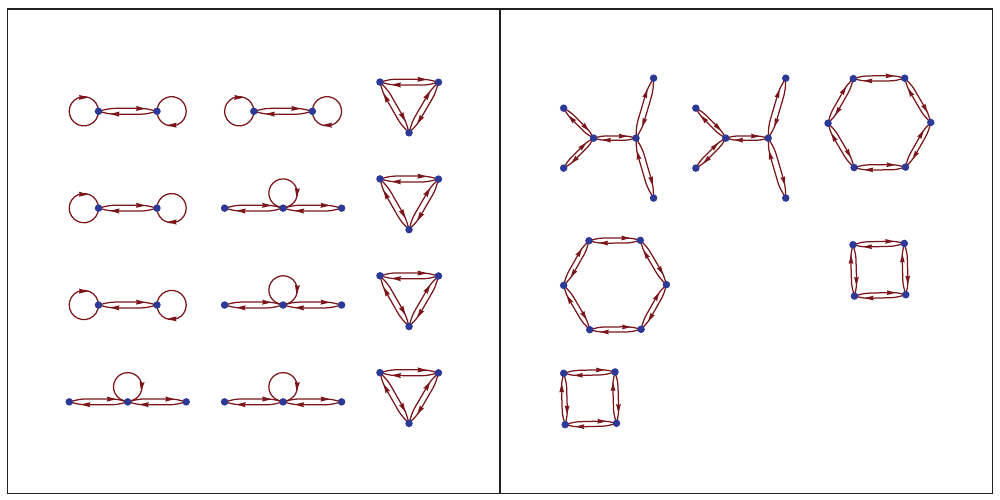}
\\
\includegraphics[scale=0.5]{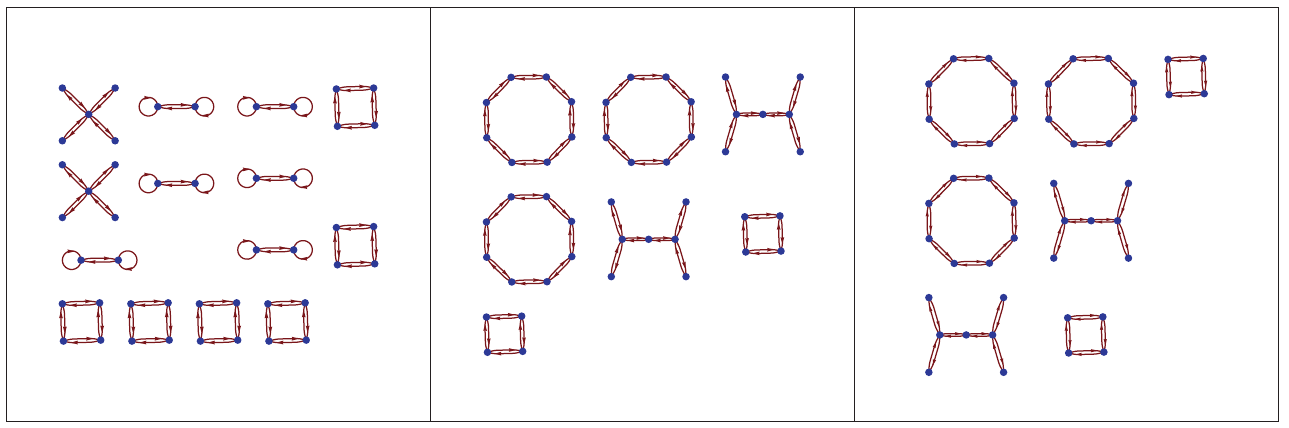} \hspace{5mm} \includegraphics[scale=0.55]{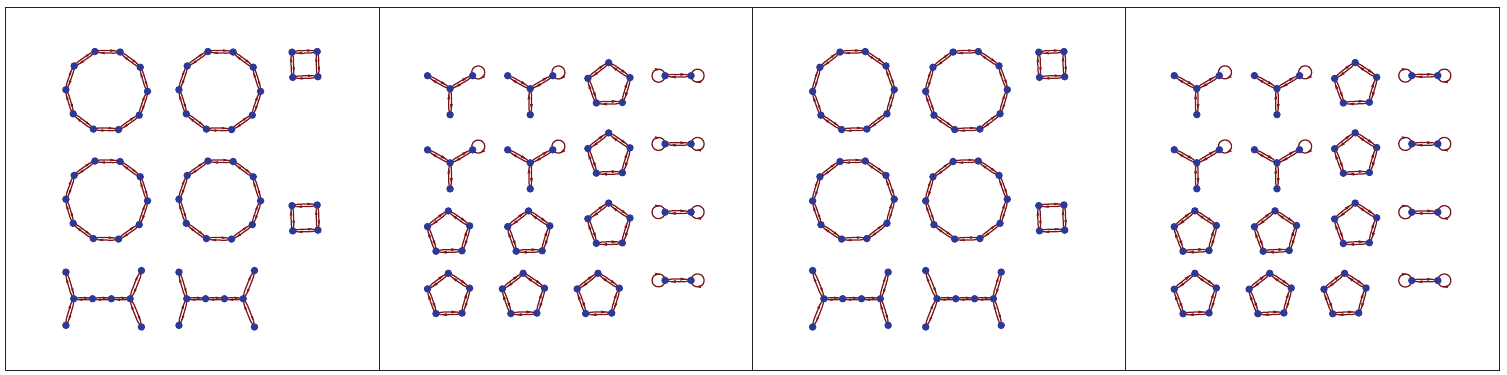}
\caption{Fusion graphs of the classical irreps of dimension $2$ for the binary dihedral
$\widehat{D}_2$, $\widehat{D}_3$, $\widehat{D}_4$, $\widehat{D}_5$.
The graphs of the faithful irreps have a~number of connected components equal to the class number
(resp.\ 5, 6, 7, 8).}\label{fig:D2D3D4D5}
\end{figure}

The smallest binary dihedral group (or quaternion group) $\hat D_2$ has only one such irrep, and it is
quaternionic, but higher $\hat D_n$'s have $2$-dimensional irreps that may be real and non faithful or
quaternionic and faithful; as it was explained, we shall call embedding representations with respect to
${\rm SU}(2)$, only those that are faithful.
One can check on Fig.~\ref{fig:D2D3D4D5} giving for binary dihedral groups all the fusion graphs
associated with $2$-dimensional
representations of the classical block of the Drinfeld double, that, as expected, only the faithful ones
have a~number of components equal to the number of classical irreps.

The binary tetrahedral group $\hat T$ has three $2$-dimensional irreps and they are faithful: only one
(that we label 4) is quaternionic, whereas those labelled 5 and 6 are complex (and conjugated).
The fusion graph associated with $N_4$, that we call embedding representation with respect to~${\rm
SU}(2)$, is the af\/f\/ine $E_6$ graph (by McKay correspondence, see below); the fusion graphs associated
with $N_5$ or $N_6$ are also connected graphs (these representations are faithful!), but they have rather
dif\/ferent features.
The binary octahedral group has also three $2$-dimensional irreps, but one ($N_3$) is real and not
faithful, the other two, $N_4$ and $N_5$, that we call embedding irreps, are both faithful and quaternionic.
The binary icosahedral group has two $2$-dimensional irreps ($N_2$, $N_3$), they are both faithful and quaternionic.
\begin{figure}[htp]\centering
\includegraphics[width=6.5cm]{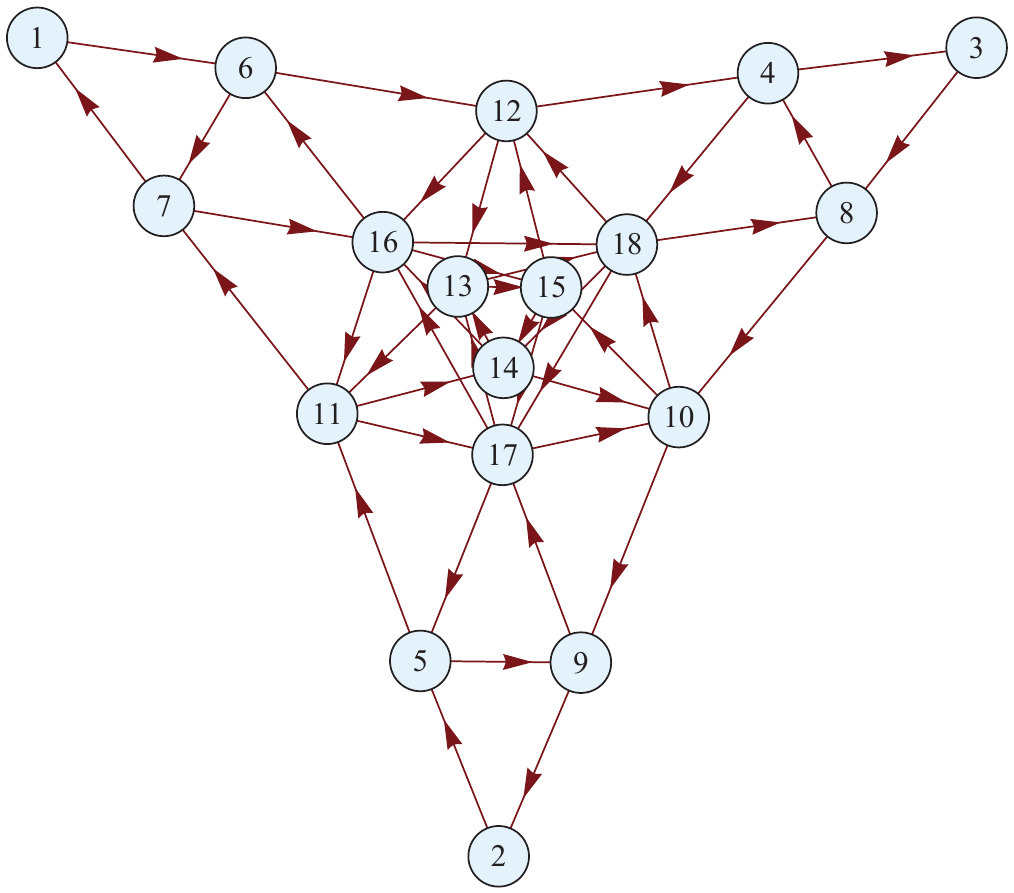}
\caption{Fusion graph $N_6$ of the subgroup $\Sigma_{168}\times\mathbb{Z}_3$.}
\label{fig:Sigma168xZ3}
\end{figure}

The case of ${\rm SU}(3)$ is a~bit more subtle because irreducible subgroups like $\Sigma_{60}$ may be
asso\-ciated with imprimitive embedding.
At this place we just illustrate on one example the importance of the faithfulness requirement: the group
$\Sigma_{168}\times\mathbb{Z}_3$ has $18$ irreps, six of them, actually three pairs of complex conjugates,
labelled $4$, $5$, $6$, $7$, $8$, $9$ are of dimension $3$, but the irreps of the pair $(4,5)$ are not faithful whereas
the two pairs $(6,7)$ and $(8,9)$ are.
As it happens the fusion graphs associated with the faithful representations (namely those labeled
$6$, $7$, $8$, $9$, $11$, $12$, $14$, $15$, $17$, $18$) are connected; this is not so for the others, in particular for the
$3$-dimensional irreps labelled $4$ and $5$.
So the natural (or embedding) irreps, with respect to ${\rm SU}(3)$, are $6$, $7$, $8$, $9$ and we may choose to
draw the fusion graph of $N_6$ for instance (see Fig.~\ref{fig:Sigma168xZ3}).
The Drinfeld double of this group {$\Sigma_{168}\times\mathbb{Z}_3$} has $288$ irreps, its f\/irst block
has $18$ irreps, as expected, that we label as for the f\/inite group itself, and again, we shall give the
fusion graph of $N_6$.
This graph is no longer connected but its number of connected components is equal to the number of blocks
(also the number of conjugacy classes, or of irreps of $G$ itself, namely $18$), see below Fig.~\ref{fig:Sigma168x3_6}.
These features are shared by $N_6$, $N_7$, $N_8$, $N_9$ but not by $N_4$ or $N_5$.

The conjugacy class determined by every central element of $G$ contains this element only, and the
centralizer of a~central element of $G$ is $G$ itself, therefore the irreps of $G$ should ap\-pear~$|Z(G)|$
times in the list of irreps of $D(G)$, where $Z(G)$ is the centre of $G$.
For example, the group $\Sigma_{36\times3}$ has four pairs of $3$-dimensional complex conjugated irreps
(every one of them can be considered as an embedding irrep), and its center is $\mathbb{Z}_3$, so we expect
that $3\times(4\times2)=24$ irreps, among the $168$ irreps of the double, will have similar properties, in
particular the same quantum dimensions (namely~$3$, since, as we know, quantum dimensions of irreps of
Drinfeld doubles are integers), and also isomorphic fusion graphs (i.e., forgetting the labeling of vertices).

\section{Drinfeld doubles (examples)}\label{section4}

\looseness=-1
In the following, we review a~certain number of f\/inite subgroups of ${\rm SU}(2)$ and ${\rm SU}(3)$,
giving for each its essential\footnote{For ``famous'' groups of relatively small order, one can retrieve
a~good amount of information from~\cite{grouppropssite}.} data, the order of the group $G$, its name in GAP
nomenclature, class number, and for its Drinfeld double $D(G)$, the rank (number of irreps), $N_c$ the
dimensions of the blocks, and quantum dimensions.
In order to shorten the often long lists of quantum dimensions, we write $n_s$ when $n$ is repeated $s$
times in a~list\footnote{The subindex $s$ therefore does not refer to an $s$-integer (all our quantum
dimensions are integers in this paper!), and $s$ does not denote a~multiplicity in the usual sense.}.
We also give the label(s) of the embedding representation(s), the fusion graph of which is then displayed.
On the connected component of the identity representation (i.e., the ``classical block") one
recognizes the corresponding fusion graph of the original group $G$.
Moreover one checks that the total number of connected components of any of these embedding fusion graphs
equals the class number of $G$, as conjectured in Section~\ref{subsection31}.

\subsection[Drinfeld doubles of f\/inite subgroups of ${\rm SU}(2)$]{Drinfeld doubles of f\/inite subgroups of $\boldsymbol{{\rm SU}(2)}$}

\subsubsection[Remarks about f\/inite subgroups of ${\rm SU}(2)$]{Remarks about f\/inite subgroups of $\boldsymbol{{\rm SU}(2)}$}
\label{fin-sbgps-SU2}

{\bf General.}
${\rm Spin}(3)={\rm SU}(2)$ is the (universal) double cover of ${\rm SO}(3)$.
$\mathbb{Z}_2$ is the center of ${\rm SU}(2)$.
With every subgroup $\Gamma$ of {\rm SO}(3) is associated its binary cover $\widehat{\Gamma}$, a~subgroup of SU(2).
Then, of course, $\Gamma\cong\widehat{\Gamma}/\mathbb{Z}_2$.
Finite subgroups of ${\rm SU}(2)$ are of the following kind: cyclic, binary dihedral, binary tetrahedral,
binary octahedral, binary icosahedral.
The fusion graphs presented below refer to the fusion matrix of the embedding representation, unless stated otherwise.

{\bf Cyclic groups.}
\begin{itemize}\itemsep=0pt
\item $\mathbb{Z}_2$ cannot be a~subgroup of $\mathbb{Z}_q$ when $q$ is odd (the order of a~subgroup should
divide the order of the group!) but it is a~subgroup of $\mathbb{Z}_q$ when $q$ is even ($q=2p$), and
$\mathbb{Z}_p=\mathbb{Z}_q/\mathbb{Z}_2$.
\item Cyclic groups $\mathbb{Z}_q$, for all $q\in\mathbb{N}$, are subgroups of ${\rm SO}(3)$ and also
subgroups of ${\rm SU}(2)$.
\item When $q$ is even ($q=2p$), we may consider the subgroup $\mathbb{Z}_q$ of ${\rm SU}(2)$ as the binary
group corresponding to the subgroup $\mathbb{Z}_p$ of ${\rm SO}(3)$ ({and} this $p$ can be even or odd).
\item When $q$ is odd, $\mathbb{Z}_q$ is a~subgroup of ${\rm SU}(2)$, but not the binary of a~subgroup of
${\rm SO}(3)$.
\item The homology or cohomology groups $H_2(\mathbb{Z}_q,\mathbb{Z})\cong
H^2(\mathbb{Z}_q,\mathbb{C}^\times)\cong H^2(\mathbb{Z}_q,U(1))$ are trivial (``the Schur multiplier is
trivial'').
Hence any Schur cover of $\mathbb{Z}_q$ is equal to itself.
Nevertheless, the cyclic group of order $2p$ can be considered as an extension, by $\mathbb{Z}_2$, of
a~cyclic group of order $p$.
{$\mathbb{Z}_{2p}$} is the binary cover of $\mathbb{Z}_{p}$ but it is not a~Schur cover of the latter.
\end{itemize}

{\bf Dihedral groups and their binary covers.}
\begin{itemize}\itemsep=0pt
\item Dihedral groups $D_n$, of order $2n$ are, for all $n\in\mathbb{N}$, subgroups of ${\rm SO}(3)$.
\item The smallest one, $D_1$, of order $2$, is isomorphic {to} $\mathbb{Z}_2$ and is usually not
considered as a~dihedral group.
\item In the context of the study of the covering ${\rm SU}(2)\to{\rm SO}(3)$, all of these groups $D_n$
can be covered by subgroups {$\widehat{D}_n$} of ${\rm SU}(2)$ of order $4n$, called binary dihedral groups
(they are also called dicyclic groups).
\item The Schur multiplier of dihedral groups $H_2(D_{n},\mathbb{Z})\cong H^2(D_{n},\mathbb{C}^\times)\cong
H^2(D_{n},U(1))$ is trivial when $n$ is odd, and is $\mathbb{Z}_2$ when $n$ is even.
Nevertheless, in both cases (even or odd) one may consider the corresponding binary dihedral groups (of
order $4n$) that are subgroups of ${\rm SU}(2)$.
\end{itemize}

{\bf The tetrahedral group $\boldsymbol{T}$ and its binary cover $\boldsymbol{\widehat{T}\cong{\rm SL}(2,3)}$.}
The tetrahedral group is $T=\widehat{T}/\mathbb{Z}_2\cong A_4$ (alternating group on four elements).

{\bf The cubic (or octahedral) group  $\boldsymbol{O}$  and its binary cover $\boldsymbol{\widehat{O}}$.}
The octahedral group is $O=\widehat{O}/\mathbb{Z}_2\cong S_4$ (symmetric group on four elements)  and its
Schur multiplier is $\mathbb{Z}_2$.  Warning: $\widehat{O}$ is {\it not} isomorphic {to} ${\rm GL}(2,3)$,
although this wrong statement can be found in the literature.
It can be realized in ${\rm GL}(2,9)$, as the matrix subgroup generated by $a=((-1,1),(0,-1))$ and
$b=((-u,-u),(-u,0))$, where $u$, obeying $u^2=-1$, is an element added to $F_3$ to generate $F_9$.
We thank~\cite{OlegCubicF9} for this information.
If $w$ generates $F_9$ (so $w^9={w}$), we take $u=w^2$.
To our knowledge, this is the smallest realization of $G$ as a~matrix group, and we used it to calculate
the Drinfeld double of the binary octahedral group.
Using GAP nomenclature, $\widehat{O}$ can be recognized as SmallGroup(48,28).

{\bf The icosahedral group $\boldsymbol{I}$ and its binary cover $\boldsymbol{\widehat{I}\cong{\rm SL}(2,5)}$.}
The icosahedral group is $I\cong A_5$ (alternating group on f\/ive elements), the smallest non-abelian
simple group, and its Schur multiplier is $\mathbb{Z}_2$.

{\bf Remarks about f\/inite subgroups of $\boldsymbol{{\rm SO}(3)}$ and $\boldsymbol{{\rm SU}(2)}$ (continuation).}
\begin{itemize}\itemsep=0pt
\item Dihedral groups $D_{n}$, of order $2n$, with $n$ odd, and cyclic groups (all of them) are the only
subgroups of ${\rm SO}(3)$ that have trivial Schur multiplier.
\item The polyhedral groups are subgroups of ${\rm SO}(3)$.
The binary polyhedral groups are subgroups of ${\rm SU}(2)$.
The so-called ``full polyhedral groups'' (that we do not use in this paper) are subgroups of O(3) and
should not be confused with the binary polyhedral groups.
Notice however that the full tetrahedral group is isomorphic to the octahedral group (both being isomorphic
to $S_4$).
\item The Schur multiplier of the exceptional polyhedral groups (tetrahedral, cube and icosahedral) are non
trivial (they are equal to $\mathbb{Z}_2$) and, for them, the binary cover can be used as a~Schur cover (however a f\/inite group may have several non isomorphic Schur covers, see the remarks in Appendix~\ref{appendixC}).
The same is true, when $n$ is even, for the dihedral groups~$D_n$.
\item All discrete f\/inite subgroups $G$ of ${\rm SU}(2)$ have trivial second cohomology
$H^2(G,\mathbb{C})=1$ and thus trivial Schur multiplier.
\end{itemize}

{\bf About the fusion graphs of the doubles.}
\begin{itemize}\itemsep=0pt
\item We shall focus our attention on the embedding irreps (or one of them if there are several) as
def\/ined at the beginning of this section and on its fusion matrix and graph.
Its label will be called ``embedding label'' in the Tables below, and its fusion graph called the embedding graph.
\item In all these embedding graphs, the connected part relative to the ``classical'' representations is
isomorphic to an af\/f\/ine Dynkin diagram of type $A_{\cdot}^{(1)}$, $D_{\cdot}^{(1)}$, $E_6^{(1)}$, $E_7^{(1)}$, $E_8^{(1)}$.
This is of course nothing else than a~manifestation of the celebrated McKay correspondence in this context~\cite{JMK}.
\item Another comment is that in Drinfeld doubles of binary groups, the classes of the identity $I$ and of
its opposite $-I$ have the same centralizer, v.i.z.\ the group itself.
To any embedding representation of the double (in the block of $I$) there is an associated irreducible in the block of~$-I$.
We found it useful to draw the two graphs of these two irreducibles together in dif\/ferent colors in
several cases, see below Figs.~\ref{fig:E6sim},~\ref{fig:edge},~\ref{fig:edgeI}.
\end{itemize}

\subsubsection{Drinfeld doubles of the (binary) cyclic subgroups}

We consider the example of $\mathbb{Z}_6$.

Order of the group: $6$

{GAP nomenclature: SmallGroup(6,2)}

Class number: $\ell=6$

Rank: {(def\/ined as the number of irreps of $D(G)$}) ${r}=36$

Numbers $N_c$ of irreps of $D(G)$ in each block = ${6,6,6,6,6,6}$

Quantum dimensions: $(1_6;1_6;1_6;1_6;1_6;1_6)$ in which we use a~shorthand notation: $p_s$ indicates that
there are $s$ irreps of dimension $p$; dif\/ferent blocks are separated by semi-colons.

$d_{\mathcal B}={{2^6}{3^6}}$

Embedding labels: 4$\oplus$6.

\begin{figure}[htp]\centering
\includegraphics[width=6.4cm]{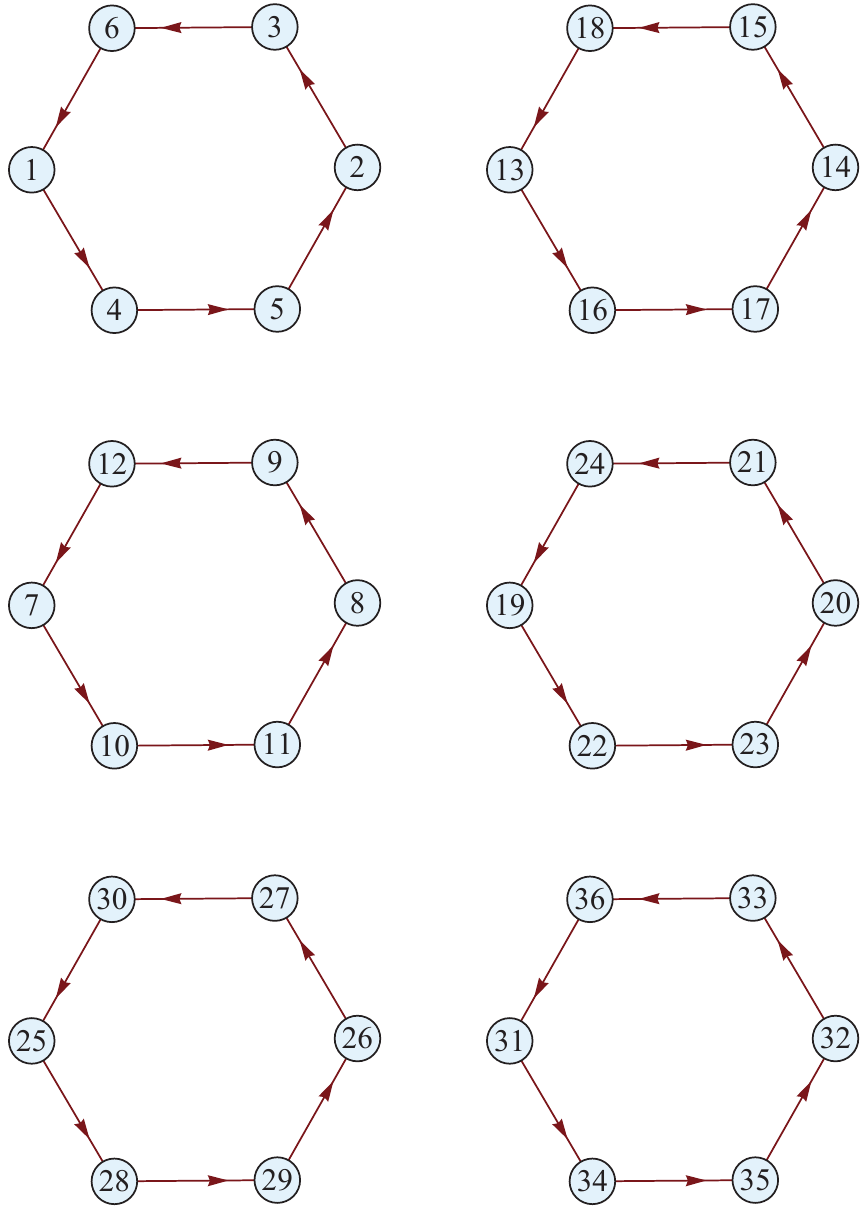}
\caption{Fusion graph $N_4$ of the Drinfeld double of the cyclic group $\mathbb{Z}_6$.}
\label{fig:Z6}
\end{figure}

\looseness=-1
As explained above, in such an abelian group in which irreps are one-dimensional, the 2-dimensional
embedding representation is the direct sum of two irreps: its is reducible as a~complex representation,
irreducible as a~real one.) See the fusion graph of $N_4$ on Fig.~\ref{fig:Z6}.
The ``embedding graph'' associated with $N_4+N_6$ looks the same, but with unoriented edges between
vertices.

Take $G=\mathbb{Z}_p$, with $p$ odd.
This is not a~binary cover, but it is nevertheless a~subgroup of~${\rm SU}(2)$.
Only the trivial representation is of real type.
All others are complex.
Observation: Only the trivial representation has non-vanishing $\Sigma$.
In particular, all complex representations of the double have vanishing $\Sigma$.
The sum rule~\eqref{sumrule} holds.

Take $G=\mathbb{Z}_p$, with $p$ even (so $G$ is a~binary cover).
There are several representations of real type, all the others being complex.
Observation: Only the trivial representation has non-vanishing $\Sigma$.
In particular, all complex representations of the double have vanishing $\Sigma$.
The sum rule~\eqref{sumrule} holds.

\subsubsection{Drinfeld doubles of the binary dihedral subgroups}

We consider the example of $\widehat{D}_5$.

Order of the group: $20$

{GAP nomenclature: SmallGroup(20,1)}

Class number: $\ell=8$

Rank: ${r}=64$

$
N_c={8,8,4,4,10,10,10,10}
$

Quantum dimensions: $(1_4,2_4;1_4,2_4;5_4;5_4;2_{10};2_{10};2_{10};2_{10})$

$
d_{\mathcal B}={{2^8}{4363^1}}
$

\begin{figure}[htp]\centering
\includegraphics[width=7.2cm]{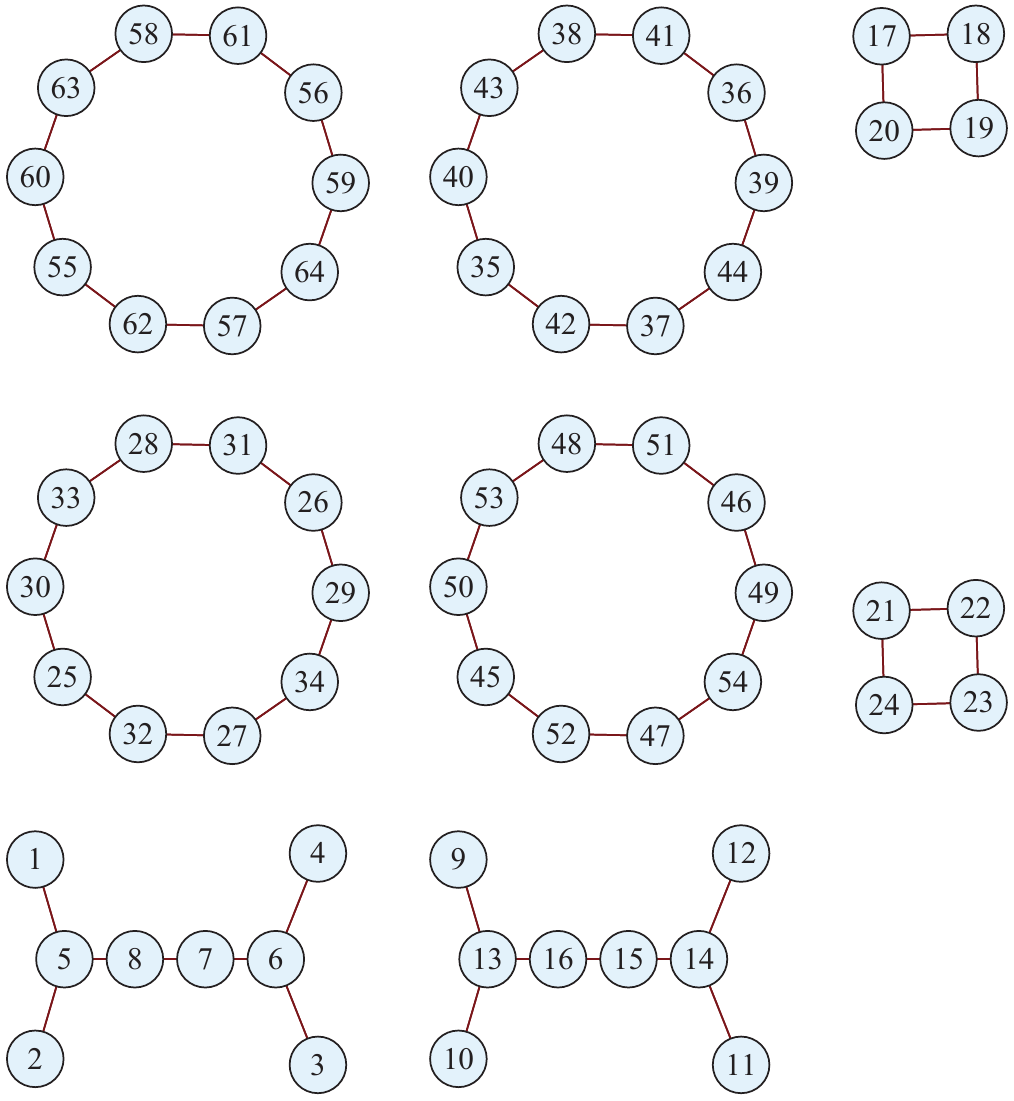}
\caption{Fundamental fusion graph of $N_5$ in the Drinfeld double of the binary dihedral
$\widehat{D}_5$.}
\label{fig:D5}
\end{figure}

{Embedding labels:} {$5,7$}.
See the fusion graph of $N_5$ on Fig.~\ref{fig:D5}.
That of $N_7$ looks very similar, up to a~permutation of labels.

The f\/irst $n+3$ irreps of the Drinfeld double of the binary dihedral group $\widehat{D}_n$ are classical
(they can be identif\/ied with the irreps of $\widehat{D}_n$).
They are of dimensions $1$ or $2$.
Their square sum is $4n$.
Among them, $n-1$ are of dimension $2$ (the others are of dimension $1$).
We draw below (Fig.~\ref{fig:D2D3D4D5}) the fusion graphs associated with these irreps of dimension~$2$.

  $\widehat{D}_2$: ${N_c=}{5,5,4,4,4}$.
Quantum dimensions: $(1_4,2_1;1_4,2_1;2_4;2_4;2_4)$.

  $\widehat{D}_3$: ${N_c=}{6,6,6,4,4,6}$.
Quantum dimensions: $(1_4,2_2;1_4,2_2;2_6;3_4;3_4;2_6)$.

  $\widehat{D}_4$: $N_c={7,7,8,4,4,8,8}$.
Quantum dimensions: $(1_4,2_3;1_4,2_3;2_8;4_4;4_4;2_8;2_8)$

  $\widehat{D}_5$: $N_c={8,8,4,4,10,10,10,10}$.
Quantum dimensions: $(1_4,2_4;1_4,2_4;5_4;5_4;2_{10};2_{10};2_{10};2_{10})$.

We analyse the case of $\widehat{D}_5$, the other tested cases are similar.
The 12 irreps labeled {3, 4, 11, 12, 17, 18, 19, 20, 21, 22, 23, 24} on Fig.~\ref{fig:D5} are complex.
The others ($64-12=52$) are self-conjugate, with 28 being real and 24 quaternionic.
The sum rule~\eqref{sumrule} holds.
Note: The sum $\Sigma_j$ vanishes for 50 irreps: the 12 irreps that are complex, but also for 38 others,
that are self-conjugate, namely all the 24 quaternionic and 14 real.
On the other hand $\Sigma_j$ does not vanish on the 14 real irreps {1, 2, 6, 8, 25, 27, 29, 31, 33, 35, 37,
39, 41, 43}.
All these vanishing $\Sigma$ follow from the existence of some unit, as in Proposition~\ref{proposition3}.
Thus there are no accidental cancellation in that case.
See Table~\ref{ssgrSU2} for a~summary.

\begin{table}[h]
\centering

\caption{Data and status of the sumrules~\eqref{sumrule} and~\eqref{sumruleS} for Drinfeld doubles
of some subgroups of ${\rm SU}(2)$.
In each box, $n \checkmark$ means the number of irreps which satisfy the sumrule in question, the sign
$\checkmark$ alone meaning ``all of them'', $n A$ gives the number of ``accidental'' vanishings, not due to
the existence of a~unit as in Proposition~\ref{proposition3}.}\label{ssgrSU2}

\vspace{1mm}

$
\noindent\renewcommand{\arraystretch}{1.2}
\begin{array}{|c||c||c|c|c|c|c|c|c|}
\hline
{\mbox{ name}} & {\eqref{sumrule}\atop{\rm before\atop
doubling}}&r&\#i,\;i\ne\bar\imath,\;\forall\,  j\atop\sum\limits_k N_{ij}^k\buildrel{?}\over=\sum\limits_k
N_{\bar\imath j}^k&{\#\;\hbox{complex}\atop\#\;\sum\limits_j S_{i j}=0}&{\#\;\hbox{quatern.}\atop\#\sum\limits_j S_{i j}=0}&{\#\;\hbox{real}\atop\#\sum\limits_j S_{i
j}=0}&\#\hbox{units}
\\
[4pt]
\hline
\hline
\mathbb{Z}_5&\checkmark&25&\checkmark&24\atop24\checkmark\;0A&0&1\atop0&25
\\[0pt]
\hline
\mathbb{Z}_6&\checkmark&36&\checkmark&32\atop32\checkmark\;0A&0&4\atop3\checkmark\;0A&36
\\[0pt]
\hline
\hline
\widehat{D}_2&\checkmark&22&\checkmark&0&8\atop8\checkmark\;0A&14\atop6\checkmark\;0A&8
\\[0pt]
\hline
\widehat{D}_3&\checkmark&32&\checkmark&12\atop12\,\checkmark\;0A&{8}\atop{8}\checkmark\;0A&12\atop6\checkmark\;0A&8
\\[2pt]
\hline
\widehat{D}_4&\checkmark&46&\checkmark&0&20\atop20\checkmark\;0A&26\atop12\checkmark\;0A&8
\\[0pt]
\hline
\widehat{D}_5&\checkmark&64&\checkmark&12\atop12\checkmark\;0A&24\atop24\checkmark\;0A&28\atop14\checkmark\;0A&8
\\[0pt]
\hline
\hline
\widehat{T}&\checkmark&42&\checkmark&32\atop32\,\checkmark\;4A&{4}\atop{4}\checkmark\;0A&6\atop2\checkmark\;2A&6
\\[2pt]
\hline
\widehat{O}&\checkmark&56&\checkmark&0&26\atop26\checkmark\;0A\,\checkmark&30\atop13\checkmark\;3A&4
\\[1pt]
\hline
\widehat{I}&\checkmark&74&\checkmark&0&36\atop36\checkmark\;0A&38\atop16\checkmark\;16A&{2}
\\[3pt]
\hline
\hline
\end{array}
$
\end{table}

\subsubsection{Drinfeld double of the binary tetrahedral}

\hspace*{5mm}Order of the group: $24$

{GAP nomenclature: SmallGroup(24,3)}.
{Alternate name: SL(2,3)}

Class number: {$\ell=7$}

Rank: ${r}=42$

$N_c=7$, $7$, $6$, $6$, $4$, $6$, $6$

Quantum dimensions: $(1_3,2_3,3_1;1_3,2_3,3_1;4_6;4_6;6_4;4_6;4_6)$

$d_{\mathcal B}={{2^5}{3^1}{13^1}{599^1}}$

\begin{figure}[htp]\centering
\includegraphics[width=8.1cm]{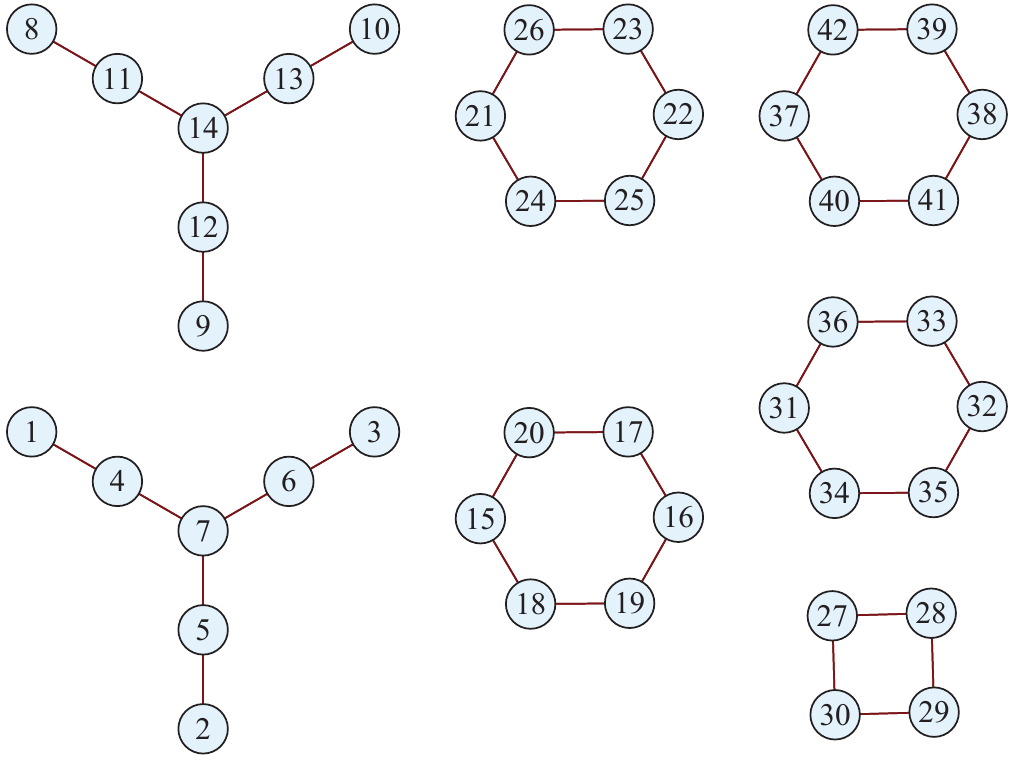}
\caption{Fusion graph $N_4$ of the Drinfeld double of the binary tetrahedral.}
\label{fig:E6}
\end{figure}

Embedding label: 4.
See Fig.~\ref{fig:E6}.

\begin{figure}[htp]\centering
\includegraphics[width=9.2cm]{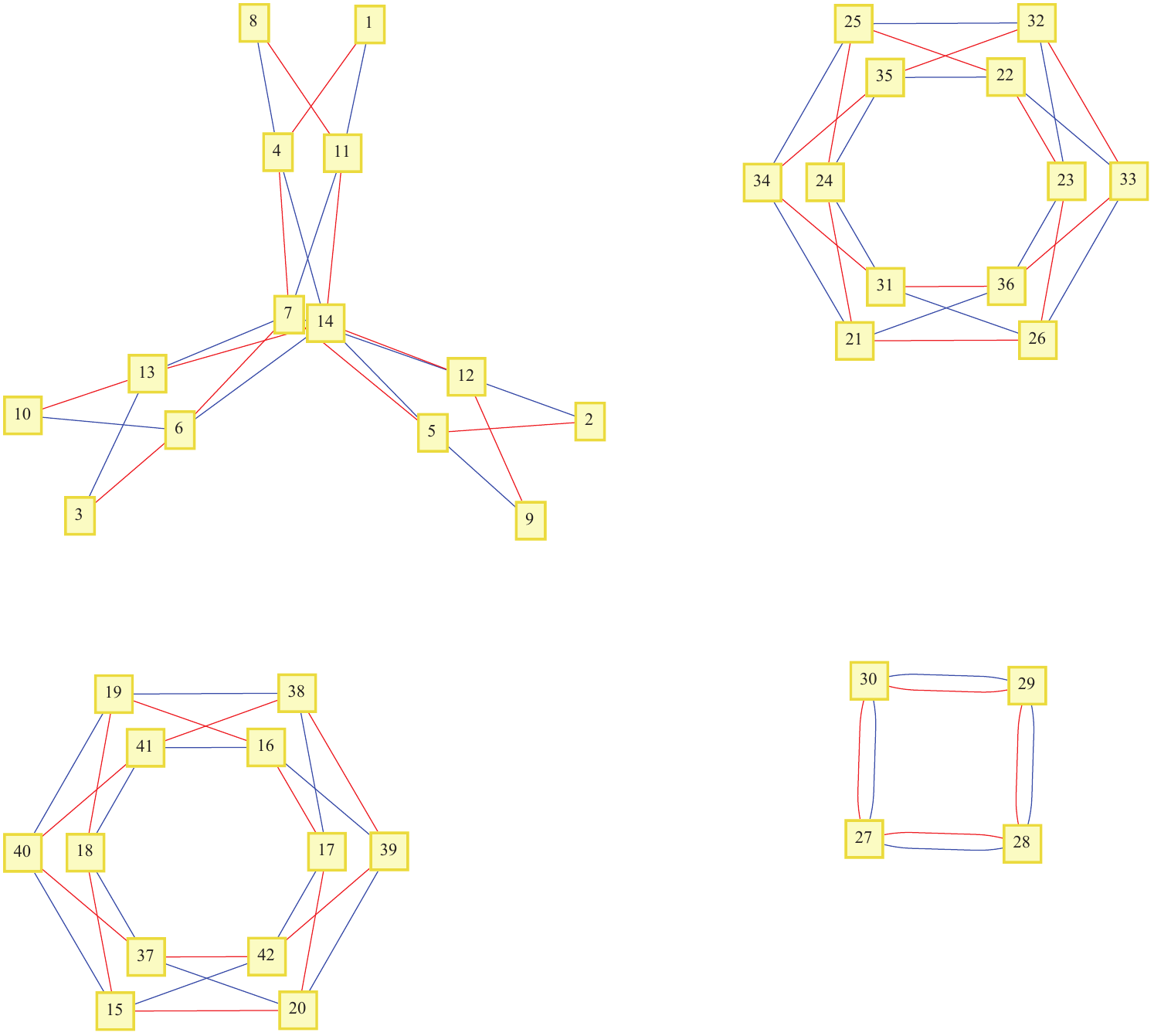}
\caption{Fusion graphs $N_4$ and $N_{11}$ of the double of the binary tetrahedral (simultaneous plot).}
\label{fig:E6sim}
\end{figure}

The graph of $N_4$ is displayed on Fig.~\ref{fig:E6}.
There is a~similar graph for the fusion matrix $N_{11}$.
See Fig.~\ref{fig:E6sim} for a~joint plot of both.

The 10 irreps labeled {1, 4, 7, 8, 11, 14, 27, 28, 29, 30} on Fig.~\ref{fig:E6} are self-conjugate.
The others ($42-10=32$) are complex.
The sum rule~\eqref{sumrule} holds.

{\bf Note.} The sum $\Sigma_j$ vanishes for 38 irreps: the 32 that are complex, but also for 6 others, including
2 real and the 4 quaternionic.
In other words $\Sigma_j$ does not vanish for the 4 real irreps~{1,~7, 27,~29}.
In 4 complex and 2 real cases, the vanishing of $\Sigma$ is accidental.

\subsubsection{Drinfeld double of the binary octahedral}

\hspace*{5mm}Order of the group: $48$

{GAP nomenclature: SmallGroup(48,28)}

Class number: {$\ell=8$}

Rank: ${r}=56$

$N_c=8$, $8$, $6$, $8$, $4$, $6$, $8$, $8$

Quantum dimensions: {$(1_2,2_3,3_2,4_1;1_2,2_3,3_2,4_1;8_6;6_8;12_4;8_6;6_8;6_8)$}

$d_{\mathcal B}={{2^7}{37447^1}}$

\begin{figure}[th!]\centering
\includegraphics[width=8.3cm]{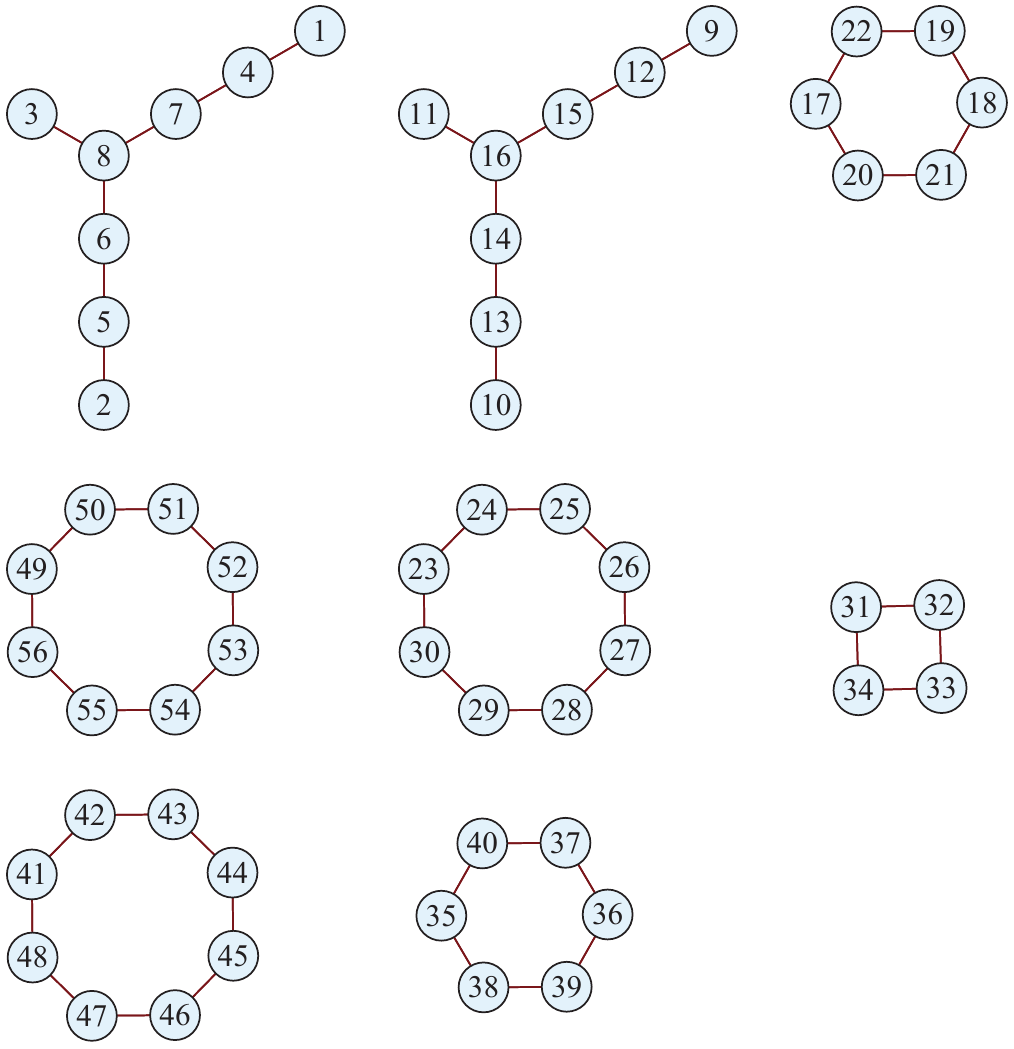}
\caption{Fusion graph $N_4$ of the Drinfeld double of the binary octahedral.}
\label{fig:E7}
\end{figure}

Embedding labels: 4, 5.
See Fig.~\ref{fig:E7}.

\begin{figure}[th!]\centering
\includegraphics[width=10.9cm]{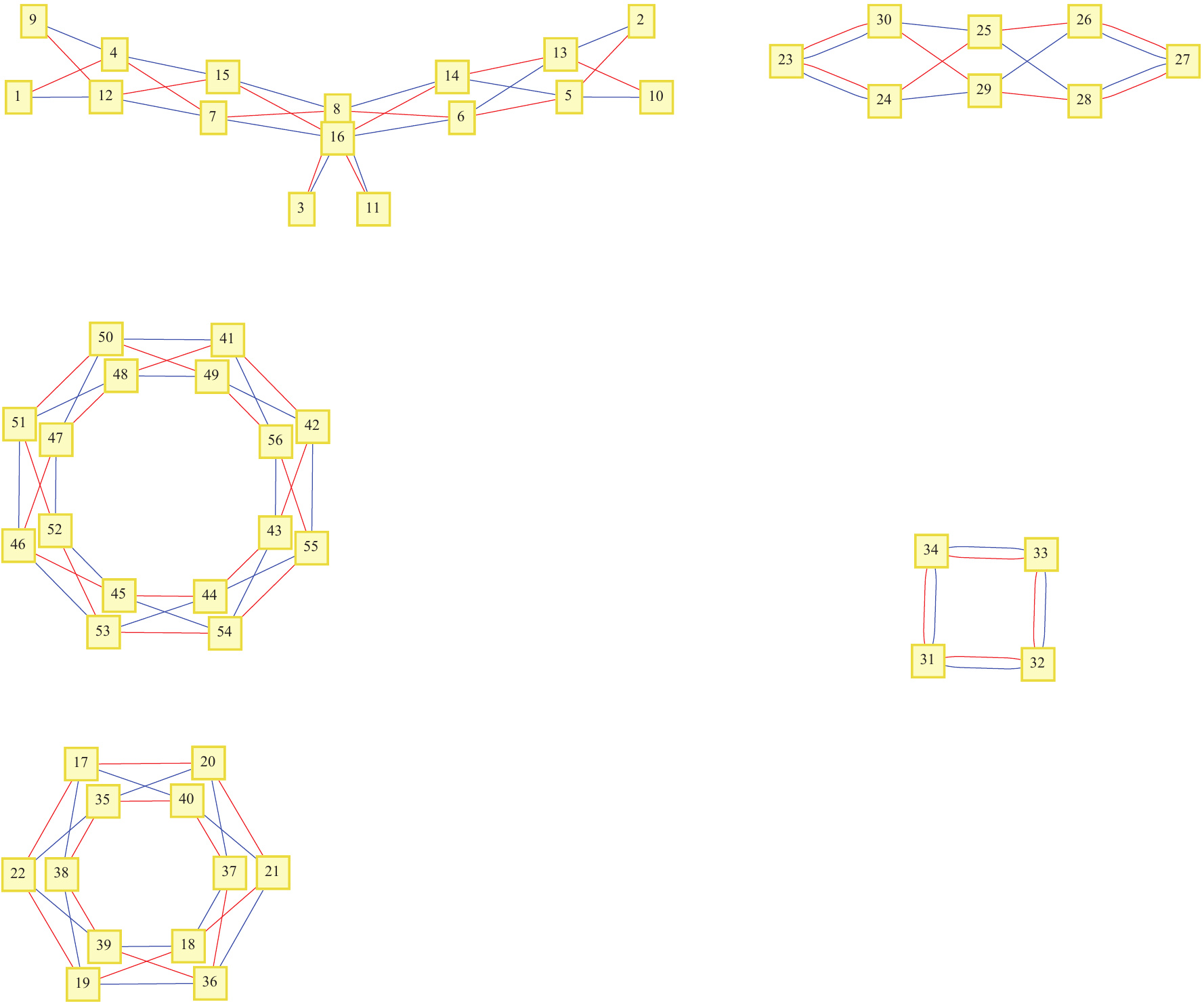}
\caption{Fusion graphs $N_4$ and $N_{12}$ of the double of the binary octahedral (simultaneous
plot).}
\label{fig:edge}
\end{figure}

There is a~similar graph for the fusion matrix $N_{12}$ (and also for $N_{5}$ and $N_{13}$).
See Fig.~\ref{fig:edge} for a~simultaneous plot of both $N_4$ and $N_{12}$.

The 56 irreps of the Drinfeld double are self-conjugate, 26 are quaternionic, 30 are real.
The sum rule~\eqref{sumrule} holds trivially.
Note: $\Sigma_j$ vanishes nevertheless for 39 irreps, all the quater\-nio\-nic~(26) and~13 real.
In three real cases, the vanishing is accidental.

\subsubsection{Drinfeld double of the binary icosahedral}

\hspace*{5mm}Order of the group: $120$

Class number: {$\ell=9$}

{GAP nomenclature: SmallGroup(120,5).} {Alternate name: SL(2,5)}

Rank: ${r}=74$

$N_c=9$, $9$, $6$, $4$, $10$, $10$, $6$, $10$, $10$

Quantum dimensions:
{$(1,2_2,3_2,4_2,5,6;1,2_2,3_2,4_2,5,6;20_6;30_4;12_{10};12_{10};20_6;12_{10};12_{10})$}

$d_{\mathcal B}={{2^5}{61^1}{89^1}{263^1}}$

\begin{figure}[th!]\centering
\includegraphics[width=8.6cm]{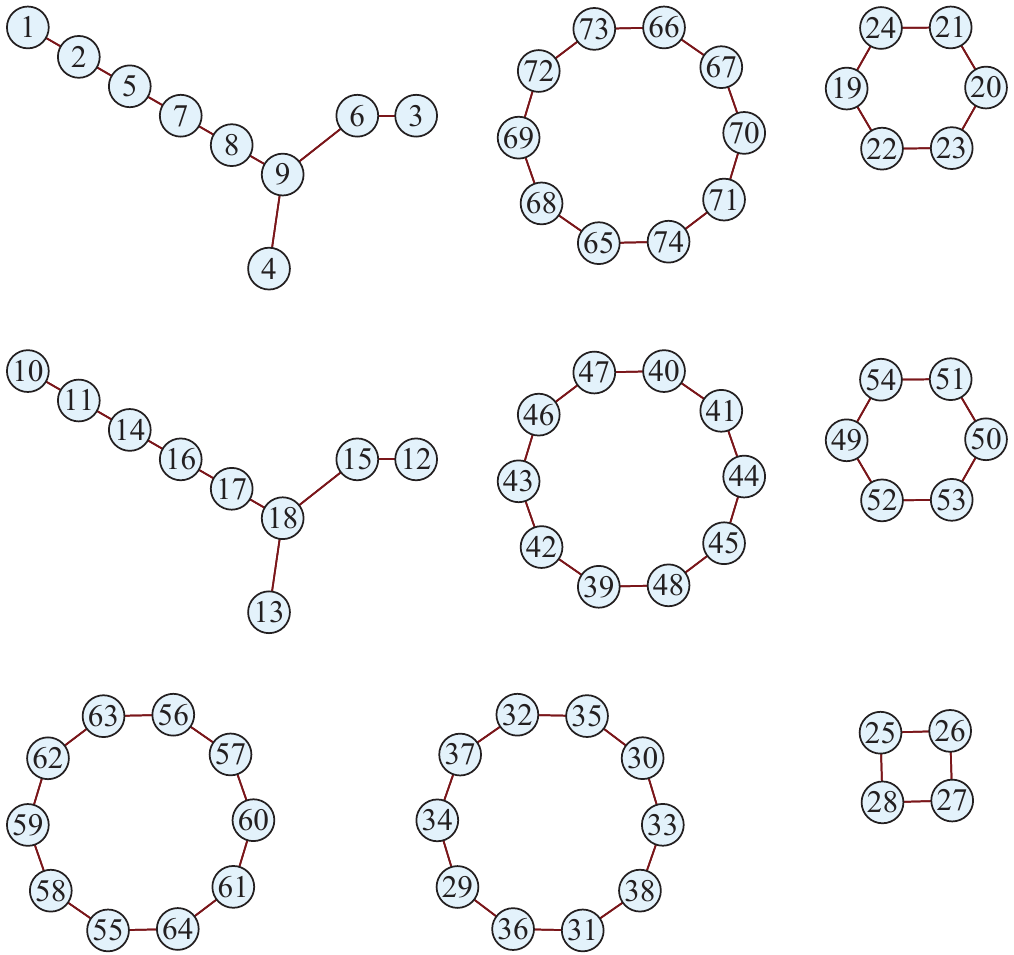}
\caption{Fusion graph $N_2$ of the Drinfeld double of the binary icosahedral.}
\label{fig:E8}
\end{figure}

\begin{figure}[th!]\centering
\includegraphics[width=8.0cm]{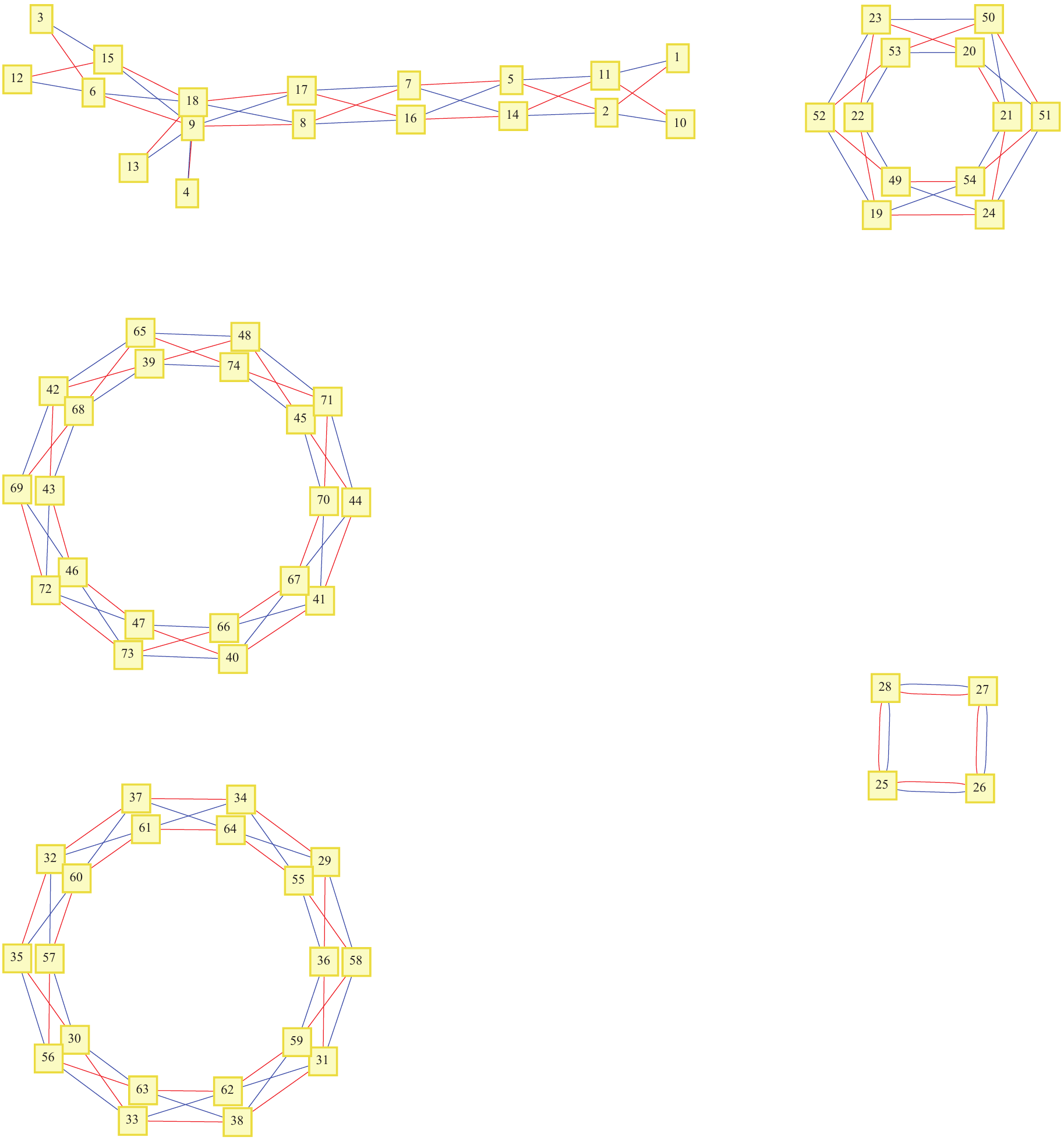}
\caption{Fusion graphs $N_2$ and $N_{11}$ of the double of the binary icosahedral (simultaneous plot).}
\label{fig:edgeI}
\end{figure}

Embedding labels: 2 and 3.
See Fig.~\ref{fig:E8}.

There is a~similar graph for the fusion matrix $N_{11}$ (and also for $N_{3}$ and $N_{12}$).

The 74 irreps of the Drinfeld double are self-conjugate: 38 are real, 36 are quaternionic.
The sum rule~\eqref{sumrule} holds trivially.
Note: $\Sigma_j$ vanishes nevertheless for 52 irreps, (16 real and all the 36 quaternionic ones).
In the 16 real cases, the vanishing is accidental.

All the data concerning sumrules~\eqref{sumrule}, \eqref{sumruleS} for doubles of subgroups of SU(2) have
been gathered in Table~\ref{ssgrSU2}.

\subsection[Drinfeld doubles of f\/inite subgroups of ${\rm SU}(3)$]{Drinfeld doubles of f\/inite subgroups of $\boldsymbol{{\rm SU}(3)}$}

\subsubsection[Remarks about f\/inite subgroups of ${\rm SU}(3)$]{Remarks about f\/inite subgroups of $\boldsymbol{{\rm SU}(3)}$}

{\bf About the classif\/ication.}
The classif\/ication of the discrete f\/inite groups of ${\rm SU}(3)$ is supposed to be
well-known, since it goes back to 1917, see~\cite{Blichfeldt}.
It has been however the object of some confusion in the more recent literature.
We recall its main lines below and proceed to the analysis of several Drinfeld doubles.

Historically, these subgroups were organized into twelve families\footnote{Or, sometimes, into only $10$
families, since $H^\star\sim H\times\mathbb{Z}_3$ and $I^\star\sim I\times\mathbb{Z}_3$ are trivially
related to $H$ and~$I$.}: four series called~$A$,~$B$, $C$,~$D$ and eight ``exceptional'' (precisely because they do
not come into series) called $E$, $F$, $G$, $H$, $I$,~$J$, and $H^\star$, $I^\star$.
However, those called $H$ and $H^\star$ are not primitive, so that the primitive exceptional are only the
six types $E$, $F$, $G$, $I$, $I^\star$, $J$ (or only f\/ive if one decides to forget about~$I^\star$).
Members of the types $A$, $B$, $C$, $D$ are never primitive: as described in~\cite{Yu}, all the~$A$, and some of the
$B$, $C$ and $D$ are subgroups of a~Lie subgroup of ${\rm SU}(3)$ isomorphic with ${\rm SU}(2)$, the other~$B$ are in ${\rm U}(2)\sim({\rm SU}(2)\times{\rm U}(1))\vert\mathbb{Z}_2$ whereas the other~$C$ and $D$, as
well as~$H$ and $H^\star$ are in a~subgroup generated by ${\rm SO}(3)\subset{\rm SU}(3)$ and the center of
${\rm SU}(3)$.
A construction of all these groups with explicit generators and relations can be found in~\cite{Roan}.
Another way of organizing these subgroups (again into $12$ families, but not quite the same) can be found
in~\cite{HuJungCai}, and a~new proposal, including only four general types, was recently made and described
in~\cite{Yu}.
A number of authors have entered the game in the past, or very recently, often with new notations and
classif\/ications, and presenting many explicit results that can be found in the following references (with
hopefully not too many omissions)~\cite{FFK,HananyEtAl:discretetorsion,Fly,Ludl1,Ludl, ParattuWinterger,Yau-Yu}.

There is no apparent consensus about the way one should classify these subgroups, not to mention the
notations to denote them (!), but everybody agrees about what they are, and in particular everybody agrees
about the list of exceptional ones.
We shall refrain from entering a~dogmatic taxonomic discussion since our purpose is mostly to discuss a~few
examples of Drinfeld doubles associated with the f\/inite subgroups of ${\rm SU}(3)$, and although we shall
consider all the exceptional cases, we shall be happy with selecting only a~few examples taken in the
inf\/inite series.
Nevertheless, for the purpose of the description, we need to present our own notations.

We organize the list of subgroups of ${\rm SU}(3)$ as follows:

With notations of the previous section, we have the subgroups $\mathbb{Z}_m\times\mathbb{Z}_n$ and
$\langle \mathbb{Z}_m,\widehat{D}_n\rangle $, $\langle \mathbb{Z}_m,\widehat{T}\rangle $, $\langle \mathbb{Z}_m,\widehat{O}\rangle $,
$\langle \mathbb{Z}_m,\widehat{I}\rangle $ whose origin can be traced back to the fact that ${\rm U}(2)\sim({\rm
U}(1)\times{\rm SU}(2))\vert\mathbb{Z}_2$ is a~subgroup of ${\rm SU}(3)$.
Here, by $\langle G_1,G_2\rangle $ we mean the group generated by the elements of the ${\rm SU}(3)$ subgroups $G_1$ and
$G_2$ (this is not, in general, a~direct product of $G_1$ and $G_2$).
The orders of the above subgroups are respectively: $m\times n$, $m\times2n$, $m\times24$, $m\times48$,
$m\times120$ if $m$ is odd, and $m\times n$, $m/2\times2n$, $m/2\times24$, $m/2\times48$, $m/2\times120$ if~$m$ is even.

Then we have the ordinary dihedral $D_n$ (of order $2n$) as well as the subgroups $\langle \mathbb{Z}_m,D_n\rangle $, of
order $m\times2n$ if $m$ or $n$ is odd, and of order $m/2\times2n$ if $m$ and $n$ are both even.

We have two series of subgroups, called $\Delta_{3n^2}$
{$=(\mathbb{Z}_n\times\mathbb{Z}_n)\rtimes\mathbb{Z}_3$}, and $\Delta_{6n^2}$
{$=(\mathbb{Z}_n\times\mathbb{Z}_n)\rtimes S_3$}, for all positive integers $n$, that appear as ${\rm
SU}(3)$ analogues of the binary dihedral groups (their orders appear as indices and the sign $\rtimes$
denotes a~semi-direct product).

The $\Delta_{3n^2}$ and $\Delta_{6n^2}$ may themselves have subgroups that are not of that kind.
For instance, for specif\/ic values of $p$ and $q$, with $p\neq q$, we have subgroups of the type
$(\mathbb{Z}_p\times\mathbb{Z}_q)\rtimes\mathbb{Z}_3$ or $(\mathbb{Z}_p\times\mathbb{Z}_q)\rtimes S_3$, but
we shall not discuss them further with the exception of Frobenius subgroups (see next entry) for the reason
that such a~digression would lie beyond the scope of our paper.
Moreover the construction of a~semi-direct product also involves a~{twisting} morphism (that is not
explicit in the previous notation) and, for this reason, it may happen that two groups built as semi-direct
products from the same components are nevertheless non-isomorphic.
The structure of the smallest ${\rm SU}(3)$ subgroups that cannot be written as direct product with cyclic
groups, that do not belong to the previous types, that are neither exceptional (see below) nor of the
Frobenius type (see next item) can be investigated from GAP.
One f\/inds $(\mathbb{Z}_9\times\mathbb{Z}_3)\rtimes\mathbb{Z}_3$,
$(\mathbb{Z}_{14}\times\mathbb{Z}_2)\rtimes\mathbb{Z}_3$, $\mathbb{Z}_{49}\rtimes\mathbb{Z}_3$,
$(\mathbb{Z}_{26}\times\mathbb{Z}_2)\rtimes\mathbb{Z}_3$,
$(\mathbb{Z}_{21}\times\mathbb{Z}_3)\rtimes\mathbb{Z}_3$,
$(\mathbb{Z}_{38}\times\mathbb{Z}_2)\rtimes\mathbb{Z}_3$, $(\mathbb{Z}_{91}\rtimes\mathbb{Z}_3)^\prime$,
$(\mathbb{Z}_{91}\rtimes\mathbb{Z}_3)^{\prime\prime}$,
$(\mathbb{Z}_{18}\times\mathbb{Z}_6)\rtimes\mathbb{Z}_3$,
$(\mathbb{Z}_{28}\times\mathbb{Z}_4)\rtimes\mathbb{Z}_3$,~$\ldots$, {see the tables displayed
in~\cite{ParattuWinterger}} or~\cite{Ludl512}.
To illustrate a~previous remark, notice that the {above} list includes two non-isomorphic subgroups of the
type $\mathbb{Z}_{91}\rtimes\mathbb{Z}_3$, recognized as SmallGroup(273,3) and SmallGroup(273,4), that
dif\/fer by the choice of the twisting morphism.

The Frobenius subgroups.
They are of the type $F_{3m}=\mathbb{Z}_m\rtimes\mathbb{Z}_3$, but $m$ should be prime of the type $6p+1$,
 i.e.,  $m=7$, $13$, $19$, $31$, $\ldots$. Their order is therefore $3m=21$, $39$, $57$, $93$, $\ldots$.
These subgroups are themselves subgroups of the $\Delta_{3n^2}$ family, but they share common features and are somehow important for us.
Indeed we checked that the property~\eqref{sumrule} fails systematically for them, and~\eqref{sumrule}, \eqref{sumruleS} fails systematically for their Drinfeld double.
We shall explicitly describe the Drinfeld double of $F_{21}$.
The latter group was recently used for particle physics phenomenological purposes in~\cite{RamondEtAl}.

The above subgroups exhaust the inf\/inite series $A$, $B$, $C$, $D$ of~\cite{Blichfeldt}, for instance the abelian
subgroups $\mathbb{Z}_m\times\mathbb{Z}_n$ correspond to the diagonal matrices (the A type), but we prefer
to describe explicitly a~subgroup by its structure, for instance such as it is given by GAP, so that
obtaining a~direct correspondence with groups def\/ined in terms of generators and relations, an
information that we did not recall anyway, is not expected to be obvious.

We are now left with the eight exceptional subgroups.
We sort them by their order and call them $\Sigma_{60}$, $\Sigma_{36\times3}$, $\Sigma_{168}$,
$\Sigma_{60}\times\mathbb{Z}_3$, $\Sigma_{72\times3}$, $\Sigma_{168}\times\mathbb{Z}_3$,
$\Sigma_{216\times3}$, $\Sigma_{360\times3}$.
They correspond respectively to the groups $H$, $E$, $I$, $H^\star$, $F$, $I^\star$, $G$, $J$, in this order, of the Blichfeldt
classif\/ication.
The structure of these groups will be recalled later (see in particular Table~\ref{infossgrSU3}).
Six of them are ternary covers of exceptional groups in ${\rm SU}(3)/\mathbb{Z}_3$ called $\Sigma_{36}$,
$\Sigma_{60}$, $\Sigma_{72}$, $\Sigma_{168}$, $\Sigma_{216}$ and $\Sigma_{360}$.
There is a~subtlety: the subgroups $\Sigma_{60}$ and $\Sigma_{168}$ of ${\rm SU}(3)/\mathbb{Z}_3$ are also
subgroups of ${\rm SU}(3)$; their ternary covers in ${\rm SU}(3)$ are isomorphic to direct products by
$\mathbb{Z}_3$ (this explains our notation); the other ternary covers are not direct products by
$\mathbb{Z}_3$.
Thus both $\Sigma_{60}$ and $\Sigma_{168}$, together with $\Sigma_{60}\times\mathbb{Z}_3$ and
$\Sigma_{168}\times\mathbb{Z}_3$, indeed appear in the f\/inal list.
Remember that only six exceptional subgroups~-- not the same ``six'' as before~-- def\/ine primitive
inclusions in ${\rm SU}(3)$, since $\Sigma_{60}$ (the icosahedral subgroup of ${\rm SO}(3)$) and
$\Sigma_{60}\times\mathbb{Z}_3$ don't.
Some people use the notation $\Sigma_p$, where $p$ is the order, to denote all these exceptional subgroups,
but this notation blurs the above distinction, and moreover it becomes ambiguous for $p=216$ since
$\Sigma_{216}$ (a subgroup of ${\rm SU}(3)/\mathbb{Z}_3$) is not isomorphic to $\Sigma_{72\times3}$ (a
subgroup of ${\rm SU}(3)$).
\begin{table}[t]
\centering
\caption{Extra information on some subgroups of SU(3).
Here, the integers $n$ always refer to cyclic groups $\mathbb{Z}_n$, so that, for instance, $2\times2$
means $\mathbb{Z}_2\times\mathbb{Z}_2$.
Unless specif\/ied otherwise, the chains of symbols in the second and penultimate columns should be left
parenthesized; for instance $3\times3\rtimes3\rtimes Q_8\rtimes3$ means
$((((\mathbb{Z}_3\times\mathbb{Z}_3)\rtimes\mathbb{Z}_3)\rtimes Q_8)\rtimes\mathbb{Z}_3)$.
Here $Q_8\cong {\rm Dic}_2\cong\widehat{D}_2$ is the quaternion group.
{$A_n$} and $S_n$ are the alternating and symmetric groups.
The ``dot'' in $3.A_6$ denotes a~triple cover of $A_6$.
{$Z(G)$} denotes the center of $G$, $G^\prime$ its commutator subgroup, and $G/G^\prime$ its abelianization.
The integer $\vert Z(G)\vert\vert G/G^\prime\vert$ gives the number of units in the fusion ring.
{${\rm Aut}(G)$} is the group of automorphisms of $G$.
{${\rm Out}(G)$} is the quotient ${\rm Aut}(G)/{\rm Inn}(G)$, where ${\rm Inn}(G)$ is the group of inner automorphisms, isomorphic
with $G/Z(G)$.
Finally $M(G)$ is the Schur multiplier of~$G$.}\label{infossgrSU3}

\vspace{1mm}

$
 \renewcommand{\arraystretch}{1.2}
\begin{array}{|@{\,\,}c@{\,\,}||@{\,\,}c@{\,\,}||@{\,\,}c@{\,\,}|@{\,\,}c@{\,\,}|@{\,\,}c@{\,\,}|@{\,\,}c@{\,\,}|@{\,\,}c@{\,\,}|}
\hline
{\mbox{name}} & \mbox{structure} &  Z(G) &  G/G^\prime & {\rm Out}(G) &  {{\rm Aut}(G)} & M(G)
\\
\hline
\hline
T=\Delta(3\times 2^2) & A_4 &1 & 3 & 2 & S_4 &2
\\
\hline
O=\Delta(6\times 2^2) &S_4 &1 & 2 & 1 & S_4 &2
\\
\hline
F_{21} & 7\rtimes 3 &1 &3 &2 & 7\rtimes 3 \rtimes 2 & 1
\\
\hline
\hline
I=\Sigma_{60} &A_5 &1 &1 &2 & S_5 & 2
\\
\hline
\Sigma_{36\times 3} &3\times3\rtimes 3\rtimes 4 &3 &4 &2 \times 2 & 3 \rtimes 3 \rtimes 8 \rtimes 2 &1
\\
\hline
\Sigma_{168} &{\rm SL}(3,2) &1 & 1& 2& {\rm SL}(3,2) \rtimes 2 & 2
\\
\hline
\Sigma_{60}\times \Bbb{Z}_3 &{\rm GL}(2,4) &3 & 3 &2\times2 & 2 \times S_5 & 2
\\
\hline
\Sigma_{72\times 3} &3\times 3\rtimes 3\rtimes Q_8 &3 &2 \times 2 &S_3 & 3 \times 3 \rtimes Q_8 \rtimes 3
\rtimes 2 &1
\\
\hline
\Sigma_{168}\times \Bbb{Z}_3 &3\times {\rm SL}(3,2) & 3&3 &2 \times 2 & 2 \times ({\rm SL}(3,2) \rtimes 2)
&2
\\
\hline
\Sigma_{216\times 3} &3\times 3\rtimes 3\rtimes Q_8\rtimes 3 & 3&3 & 6& 3 \times (3 \times 3 \rtimes Q_8
\rtimes 3 \rtimes 2) &1
\\
\hline
\Sigma_{360\times 3} &3.
A_6 & 3&1 &2\times 2 &A_6\rtimes 2 \rtimes 2 & 2
\\
\hline
\end{array}
$
\end{table}

{\bf Miscellaneous remarks.} In the literature, one can f\/ind several families of generators for the
above groups; very often these generators are given as elements of ${\rm GL}(3)$, of ${\rm SL}(3)$, or of~${\rm U}(3)$, not of~${\rm SU}(3)$.
Such f\/indings do not imply any contradiction with the above classif\/ication since the latter is only
given up to isomorphism.

Because of the embeddings of Lie groups ${\rm SU}(2)\subset{\rm SU}(3)$ and ${\rm SO}(3)\subset{\rm
SU}(3)$, all the f\/inite subgroups of ${\rm SU}(2)$ are f\/inite subgroups of ${\rm SU}(3)$ and all the
f\/inite subgroups of ${\rm SO}(3)$ are subgroups of ${\rm SU}(3)$ as well.
In particular all polyhedral groups and all binary polyhedral groups are subgroups of ${\rm SU}(3)$.
Using the fact that $T\cong\Delta(3\times2^2)$, $O\cong\Delta(6\times2^2)$, {and ${I}\cong
A_5\cong\Sigma_{60}$}, the reader can recognize all of them in the above list.

Several calculations whose results are given below rely on some group theoretical information (for instance
the determination of conjugacy classes and centralisers) that was taken from the {GAP} smallgroup library.

We shall now review a~certain number of f\/inite subgroups of ${\rm SU}(3)$ and their Drinfeld double.
Like in the case of SU(2), we list for each of them a~certain of data, and display one of their embedding
fusion graphs.
These graphs become fairly involved for large subgroups, and we label their vertices only for the smallest
groups, while for the larger ones, only the connected component of the identity representation is labelled.
Data on the way sum rules~\eqref{sumrule} and~\eqref{sumruleS} are or are not satisf\/ied, and the numbers
of ``accidental'' vanishings, are gathered in Table~\ref{ssgrSU3}.

\subsubsection[Drinfeld double of $F_{21}=\mathbb{Z}_7\rtimes\mathbb{Z}_3$]{Drinfeld double of $\boldsymbol{F_{21}=\mathbb{Z}_7\rtimes\mathbb{Z}_3}$}

\hspace*{5mm}Order of the group: $21$

GAP nomenclature: $F_{21}={\rm SmallGroup}(21,1)$

Class number: $\ell=5$

Classical dimensions: {1, 1, 1, 3, 3}

Rank: ${r}=25$

$N_c=5$, $3$, $3$, $7$, $7$

Quantum dimensions: $(1_3,3_2;7_3;7_3;3_7;3_7)$

$d_{\mathcal B}={{5^1}{11^1}{23^1}{137^1}}$

\begin{figure}[th!]\centering
\includegraphics[width=7.2cm]{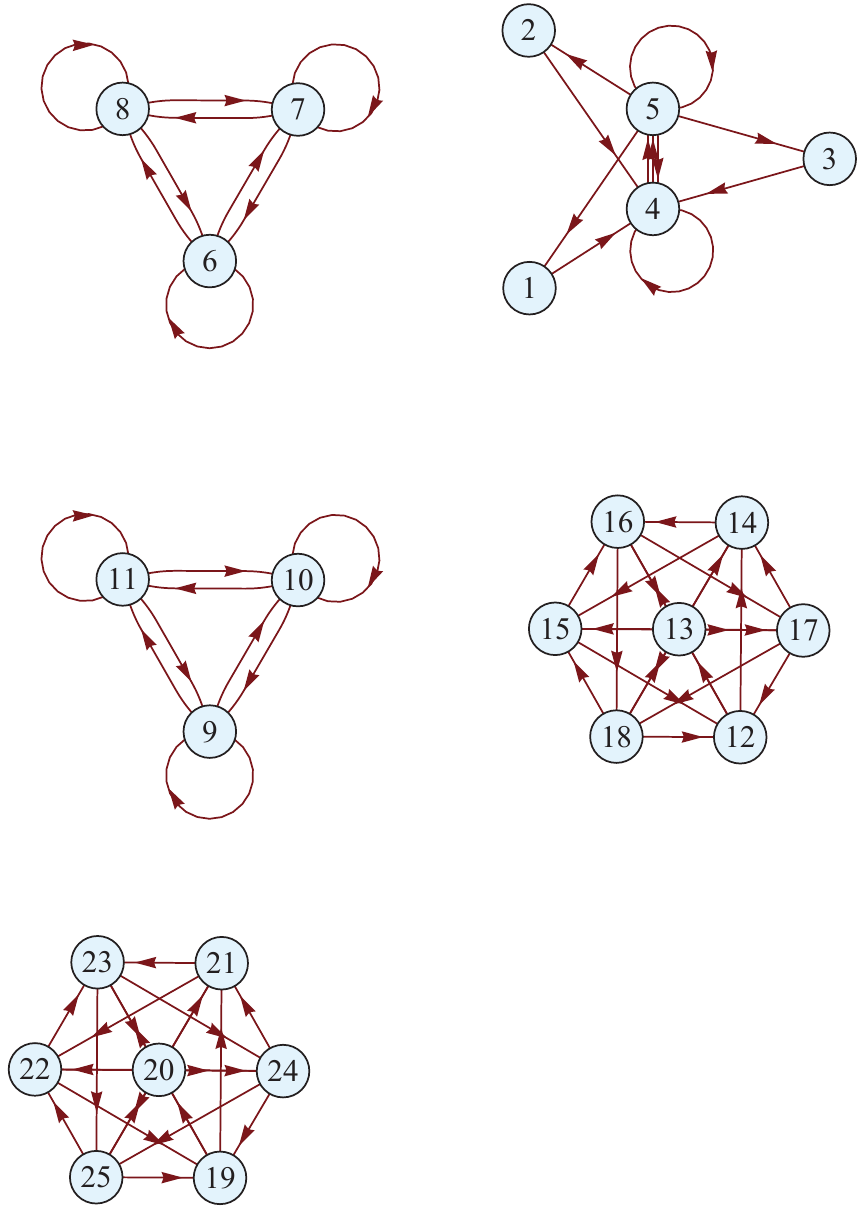}
\vspace{-1mm}
\caption{Fusion graph $N_4$ of the Drinfeld double of the group $F_{21}$.}
\label{fig:F21}
\end{figure}

Embedding labels: 4 and 5; see Fig.~\ref{fig:F21} for the fusion graph of $N_4$.

\subsubsection[Drinfeld double of $\Sigma_{60}$]{Drinfeld double of $\boldsymbol{\Sigma_{60}}$}

Remember that $\Sigma_{60}$ is both a~subgroup of ${\rm SU}(3)/\mathbb{Z}_3$ and a~subgroup of ${\rm SU}(3)$.
Also, recall that this group is isomorphic to the icosahedron group $I$ of Section~\ref{fin-sbgps-SU2}.
Thus the 5 classical irreps of its double identify with the zero-``bi-ality'' irreps of the {\it binary} icosahedron group $\widehat{I}$.

Order of the group: $60$

GAP nomenclature: ${\rm SmallGroup}(60,5)$.
Alternate names: $A_5$, ${\rm SL}(2,4)$, I (Icosahedral).

Class number: $\ell=5$

Classical dimensions: {1, 3, 3, 4, 5}

Rank: ${r}=22$

$N_c={5,4,3,5,5}$

Quantum dimensions: $(1,3_2,4,5;15_4;20_3;12_5;12_5)$

$d_{\mathcal B}={{2^1}{5^1}{11^1}{10853^1}}$

\begin{figure}[th!]\centering
\includegraphics[width=7.2cm]{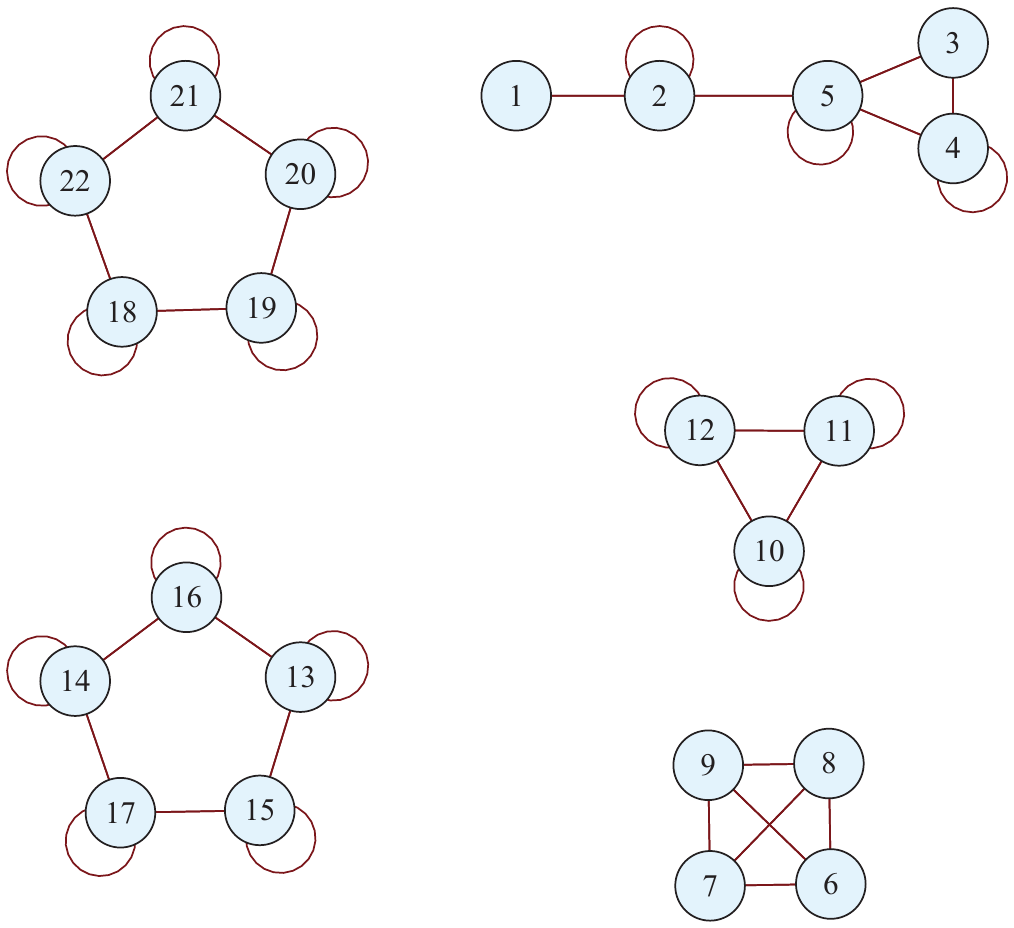}
\vspace{-1mm}
\caption{Fusion graph $N_2$ of the Drinfeld double of the group $\Sigma_{60}$.}
\label{fig:Sigma60}
\end{figure}

Embedding labels: 2 and 3; see Fig.~\ref{fig:Sigma60} for the fusion graph of $N_2$.

All the representations of $\Sigma(60)$ and of its double are real.

\subsubsection[Drinfeld double of $\Sigma_{36\times3}$]{Drinfeld double of $\boldsymbol{\Sigma_{36\times3}}$}

\hspace*{5mm}Order of the group: $108$

GAP nomenclature: $\Sigma_{36\times3}={\rm SmallGroup}(108,15)$

Class number: $\ell=14$

Classical dimensions: ${1,1,1,1,3,3,3,3,3,3,3,3,4,4}$

Rank: ${{r}}=168$

$N_c=14$, $12$, $14$, $14$, $9$, $9$, $12$, $12$, $12$, $12$, $12$, $12$, $12$, $12$

Quantum dimensions:~$(1_4,3_8,4_2;9_{12};1_4,3_8,4_2;1_4,3_8,4_2;12_{9};12_{9};9_{12};9_{12};9_{12};9_{12};9_{12};$

\hspace*{38mm}$9_{12};9_{12};9_{12})$

$d_{\mathcal B}={{2^3}{3^5}{124477^1}}$

\begin{figure}[htp]\centering
\includegraphics[width=4.7cm]{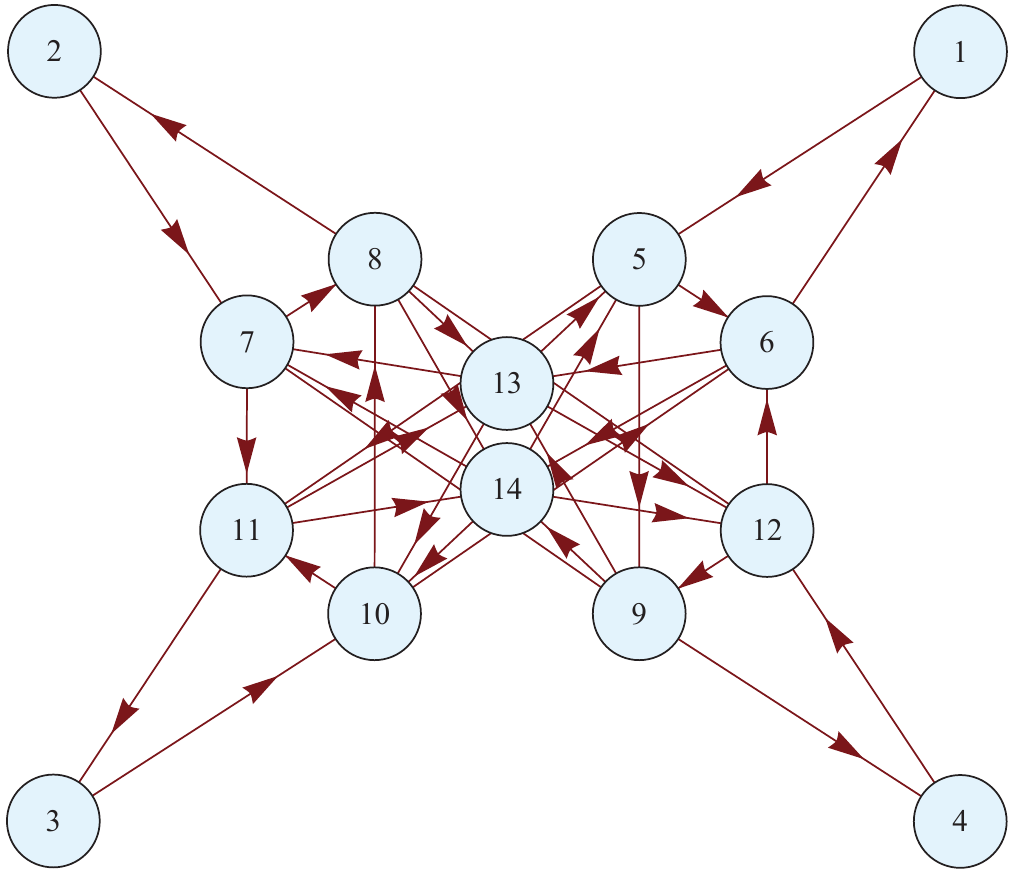}

\includegraphics[width=14.5cm]{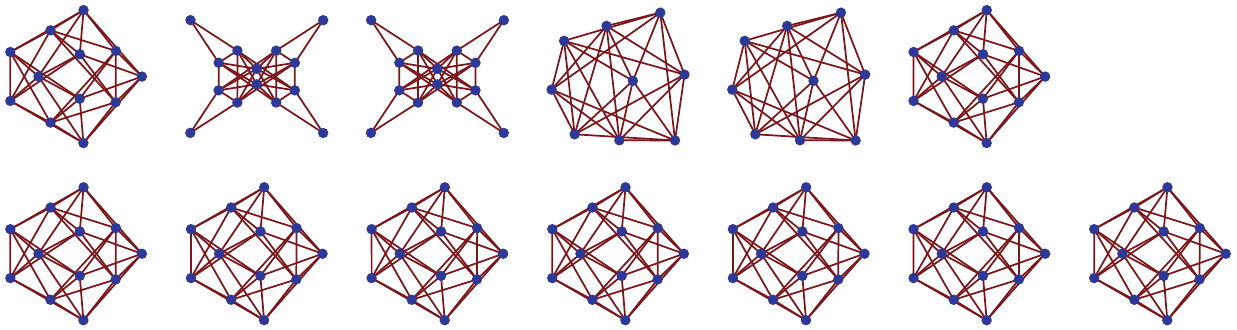}

\caption{Fusion graph $N_5$ of the Drinfeld double of the group {$\Sigma_{36\times3}$}.
Only the f\/irst connected component has been displayed with the labels of the vertices (irreps).}
\label{fig:Sigma108}
\end{figure}

Embedding labels: 5 and 6;  see Fig.~\ref{fig:Sigma108} for the fusion graph $N_5$.

For illustration, let us analyze in some details the fusion graph of the quantum double of the group
$G=\Sigma_{36\times3}$, see Fig.~\ref{fig:Sigma108}.
The class number of $G$ is $14$.
{$G$} has nine $3$-dimensional embedding representations labelled from $5$ to $12$.
They all give fusion graphs sharing the same overall features.
We have chosen to display $N_5$.
The fusion graph of $G$ itself (the ``classical graph'') appears on the top.
It is itself connected (it would not be so if we had chosen for instance $N_{13}$ or $N_{14}$, which
correspond to $4$-dimensional non faithful representations).
It has $14$~vertices.
The fusion graph of the Drinfeld double~$D(G)$ has $14$ connected components.
As the center of~$G$ is~$\mathbb{Z}_3$, the classical graph appears three times in the fusion graph of~$D(G)$.
The group~$G$ has $14$ conjugacy classes, but their stabilizers fall into only three types (up to
isomorphisms): $G$~itself for the center ($3$ times), the cyclic group $\mathbb{Z}_3\times\mathbb{Z}_3$,
that appears twice, and the cyclic group $\mathbb{Z}_{12}$, that appears nine times.
The corresponding three kinds of connected components appear on Fig.~\ref{fig:Sigma108}.
Apart from~$G$ itself, these stabilizers are abelian, so that the number of their irreps (number of
vertices of the corresponding connected components) are given by their order, respectively~$9$ and~$12$.
The unity of the fusion ring (trivial representation) is labelled~$1$.
This ring has twelve units.
Four ($1,2,3,4$), the ``classical units'', appear as the endpoints of the classical graph.
The other eight ($4+4$) are the corresponding endpoints on the two other copies of $G$.
The units $1$, $2$ are of real type, the units~$3$,~$4$ are of complex type (and actually conjugated).
The eight embedding representations $5\ldots12$ are connected to the classical units.
They appear in conjugate pairs $(5,6)$, $(7,8)$ connected respectively to $1$ and $2$, and $(9,10)$, $(11,12)$
connected to $3$ and $4$.
Notice that a~conjugated pair is attached to the same unit when this unit is real but to complex conjugate
units when the unit is complex.
Of course, the edges of the graph are oriented since $5$ is not equivalent to its complex conjugate; the
fusion graph $N_6$ can be obtained from the given graph, $N_5$, by reversing the arrows.
The action of units clearly induces geometrical symmetries on the given fusion graph, nevertheless one
expects that the fusion graphs associated with $5$, $6$, $7$, $8$ on the one hand, or with $9$, $10$, $11$, $12$ on the
other, or with any of the corresponding vertices belonging to the three copies of the classical graph,
although sharing the same overall features, will look slightly dif\/ferent, and the reader can check (for
instance by drawing the graph $N_9$) that it is indeed so.
(As a~side remark, the classical graph of $\Sigma_{36\times3}$, drawn dif\/ferently
(see~\cite[Fig.~14]{DiFrancescoZuber}) leads after amputation of some vertices and edges to the fusion graph ${\mathcal E^{(8)}}$, the
star-shaped exceptional module of ${\rm SU}(3)$ at level~$5$).
The exponent of~$G$~\cite{CGR:modulardata} is $m=12$, it is equal to the order of the modular matrix~$T$,
like for all Drinfeld doubles, and the entries of~$S$ and~$T$ (these are $168\times168$ matrices with
entries labelled by the vertices of the Fig.~\ref{fig:Sigma108}) lie in the cyclotomic f\/ield $\mathbb{Q}(\xi)$ where $\xi=\exp(2i\pi/m)$.
We did not discuss Galois automorphisms in this paper, but let us mention nevertheless that there is also
a~Galois group acting by permutation on vertices, it is isomorphic to the multiplicative group~$\mathbb{Z}_{m}^\times$ of integers coprime to~$m$,~\cite{CGR:modulardata, Gannon:modular}.

\subsubsection[Drinfeld double of $\Sigma_{168}$]{Drinfeld double of $\boldsymbol{\Sigma_{168}}$}

Remember that $\Sigma_{168}$ is both a~subgroup of ${\rm SU}(3)/\mathbb{Z}_3$ and a~subgroup of ${\rm SU}(3)$.
The group $\Sigma_{168}$ is the second smallest simple non-abelian group (the smallest being the usual
icosahedral group ${I}\cong A_5$ of course!).
It is {often} called the Klein group, or the smallest Hurwitz group.

GAP nomenclature: $\Sigma_{168}={\rm SmallGroup}(168,42)$.

Alternate names: ${\rm SL}(3,2)\cong{\rm PSL}(3,2)\cong{\rm GL}(3,2)\cong{\rm PSL}(2,7)$.

Order of the group: $168$

Class number: $\ell=6$

Classical dimensions: {1, 3, 3, 6, 7, 8}

Rank: ${r}=32$

$N_c=6$, $5$, $3$, $4$, $7$, $7$

Quantum dimensions: $(1,3_2,6,7,8;21_4,42;56_3;42_4;24_7;24_7)$

$d_{\mathcal B}={{2^2}\,{4126561^1}}$

\begin{figure}[th!]\centering
\includegraphics[width=11.0cm]{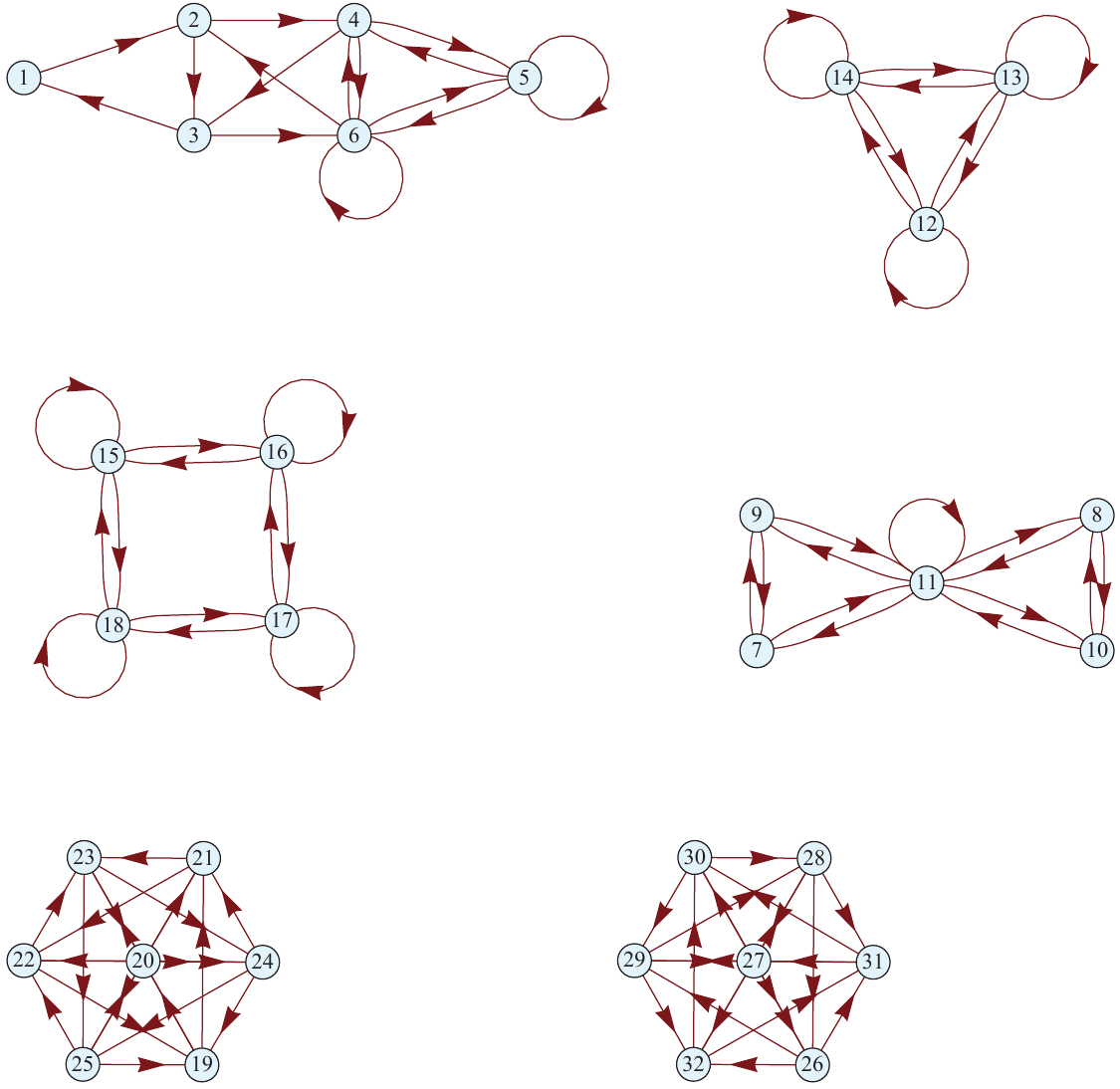}
\caption{Fusion graph $N_2$ of the Drinfeld double of the Hurwitz group $\Sigma_{168}$.}
\label{fig:Sigma168}
\end{figure}

Embedding labels: 2 and 3.
See Fig.~\ref{fig:Sigma168} for the fusion graph $N_2$.
The other embedding fusion graph $N_3$ is obtained from $N_2$~\eqref{fig:Sigma168} by reversing the arrows.

For illustration, we shall give explicitly the $S$ matrix of this Drinfeld double in Appendix~\ref{appendixD}.

\subsubsection[Drinfeld double of $\Sigma_{60}\times\mathbb{Z}_3$]{Drinfeld double of $\boldsymbol{\Sigma_{60}\times\mathbb{Z}_3}$}

\hspace*{5mm}Order of the group: $180$

GAP nomenclature: $\Sigma_{60}\times\mathbb{Z}_3={\rm SmallGroup}(180,19)$.

Alternate names: ${\rm GL}(2,4)$.

Class number: $\ell=15$

Classical dimensions: {1, 1, 1, 3, 3, 3, 3, 3, 3, 4, 4, 4, 5, 5, 5}

{Rank: ${r}=198$}

$N_c=15$, $12$, $15$, $15$, $9$, $9$, $9$, $15$, $15$, $12$, $12$, $15$, $15$, $15$, $15$

Quantum dimensions: $(1_3,3_6,4_3,5_3;15_{12};1_3,3_6,4_3,5_3;1_3,3_6,4_3,5_3;20_{9};20_{9};20_{9};12_{15};12_{15};$

\hspace*{38mm}$15_{12};15_{12};12_{15};12_{15};12_{15};12_{15})$

$d_{\mathcal B}={{2^1}{3^6}{5^1}{11^1}{10853^1}}$

Embedding labels: {6 and its conjugate 9, or 7 and its conjugate 8}

\begin{figure}[th!]\centering
\includegraphics[width=4.5cm]{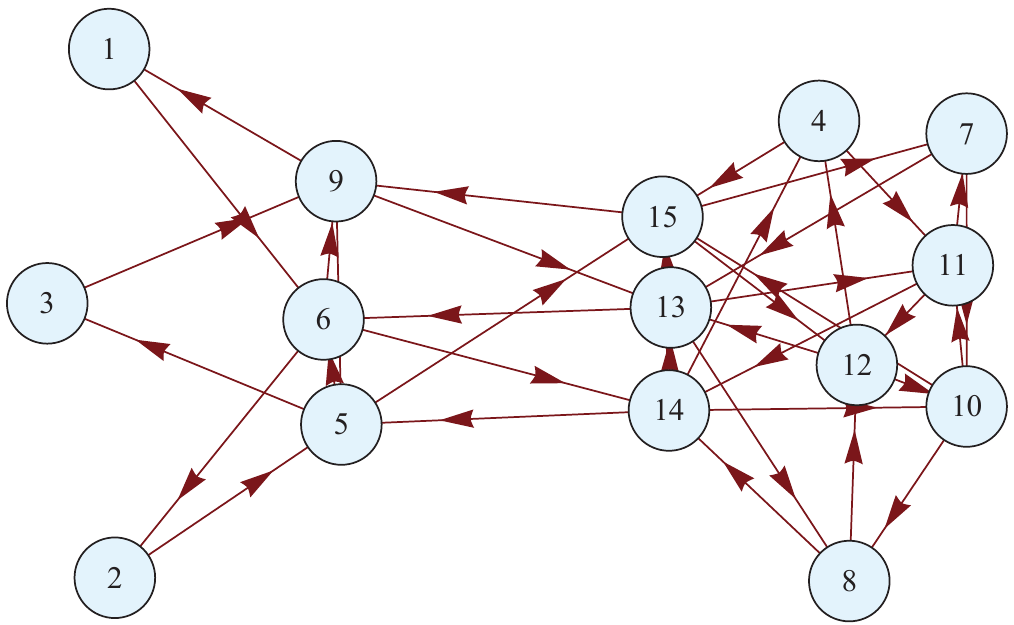}

\includegraphics[width=14.5cm]{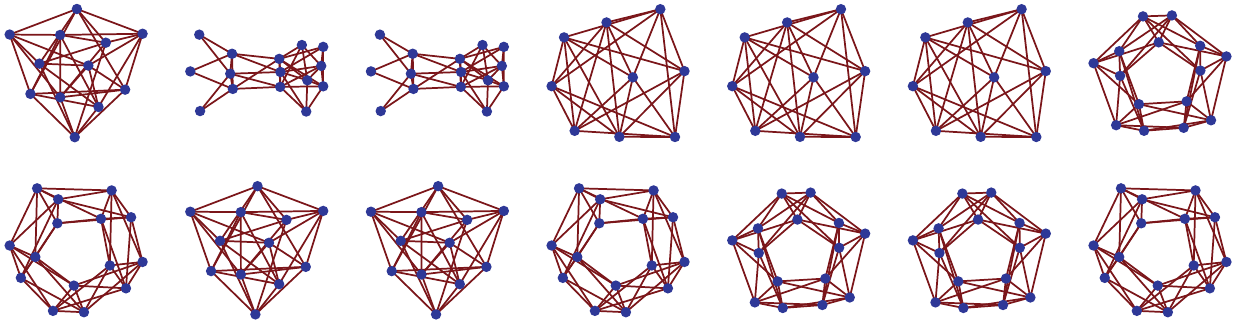}
\caption{Fusion graph $N_6$ of the Drinfeld double of the group $\Sigma_{60}\times\mathbb{Z}_3$.}
\label{fig:Sigma60xZ3}
\end{figure}

See Fig.~\ref{fig:Sigma60xZ3} for the fusion graph of {$N_6$}.

\subsubsection[Drinfeld double of $\Sigma_{72\times3}$]{Drinfeld double of $\boldsymbol{\Sigma_{72\times3}}$}

\hspace*{5mm}Order of the group: $216$

GAP nomenclature: $\Sigma_{72\times3}={\rm SmallGroup}(216,88)$

Class number: $\ell=16$

Classical dimensions: {1, 1, 1, 1, 2, 3, 3, 3, 3, 3, 3, 3, 3, 6, 6, 8}

Rank: $r=210$

$N_c=16$, $15$, $16$, $16$, $9$, $12$, $12$, $12$, $15$, $15$, $12$, $12$, $12$, $12$, $12$, $12$

Quantum dimensions:
$(1_4,2,3_8,6_2,8;9_{12},18_3;1_4,2,3_8,6_2,8;1_4,2,3_8,6_2,8;24_9;$

\hspace*{38mm}$18_{12};18_{12};18_{12};9_{12},18_{3};9_{12},18_{3};18_{12};18_{12};18_{12};18_{12};18_{12};18_{12})$

$d_{\mathcal B}={{2^2}{3^3}{23^1}{59^1}{8941^1}}$

Embedding labels: 6, \dots, 13.

\begin{figure}[th!]\centering
\includegraphics[width=5.5cm]{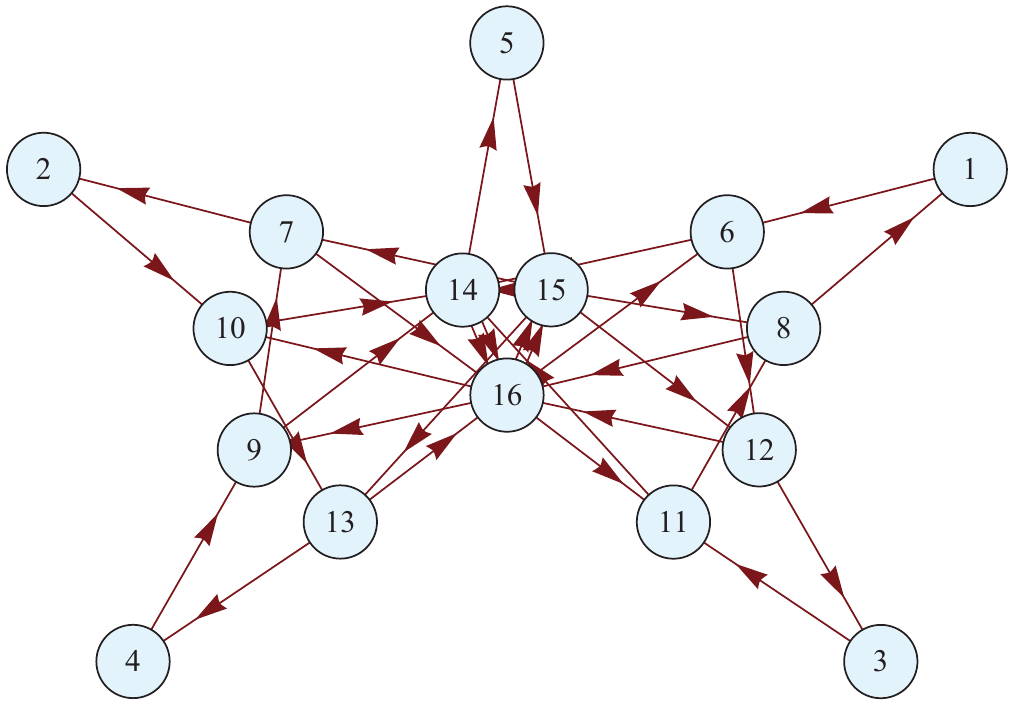}

\includegraphics[width=14.2cm]{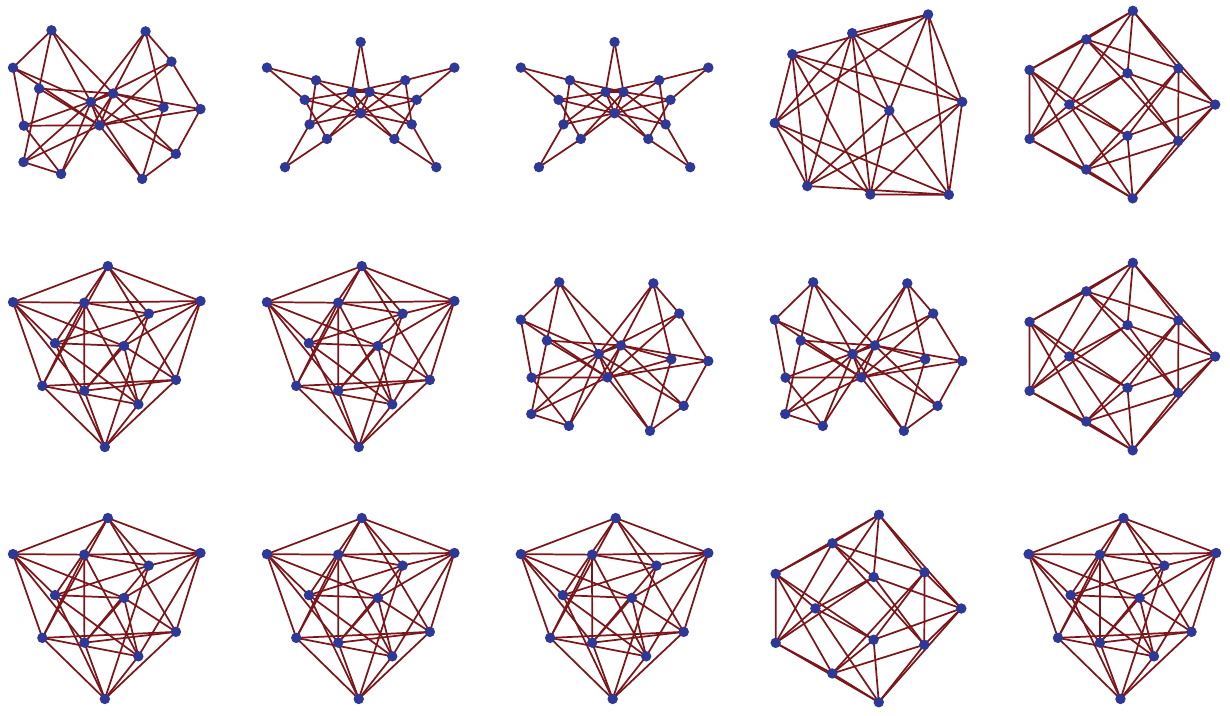}
\caption{Fusion graph $N_{{6}}$ of the Drinfeld double of the group $\Sigma_{72\times3}$.}
\end{figure}

\subsubsection[Drinfeld double of $\Sigma_{168}\times\mathbb{Z}_3$]{Drinfeld double of $\boldsymbol{\Sigma_{168}\times\mathbb{Z}_3}$}

\hspace*{5mm}Order of the group: 504

GAP nomenclature: $\Sigma_{168}\times\mathbb{Z}_3={\rm SmallGroup}(504,157)$.

Class number: $\ell=18$

Classical dimensions: $1$, $1$, $1$, $3$, $3$, $3$, $3$, $3$, $3$, $6$, $6$, $6$, $7$, $7$, $7$, $8$, $8$, $8$

Rank: $r=288$

$N_c=18$, $15$, $18$, $18$, $9$, $9$, $9$, $12$, $15$, $15$, $21$, $21$, $12$, $12$, $21$, $21$, $21$, $21$

Quantum dimensions: $(1_3,3_6,6_3,7_3,8_3;21_{12},42_3;1_3,3_6,6_3,7_3,8_3;1_3,3_6,6_3,7_3,8_3;56_9;56_9;$

\hspace*{38mm}$42_{12};21_{12},42_3;21_{12},42_3;24_{21};24_{21};42_{12};42_{12};24_{21};24_{21};24_{21};24_{21})$

$d_{\mathcal B}={{2^2}{3^6}{4126561^1}}$

\begin{figure}[htp]\centering
\includegraphics[width=4.5cm]{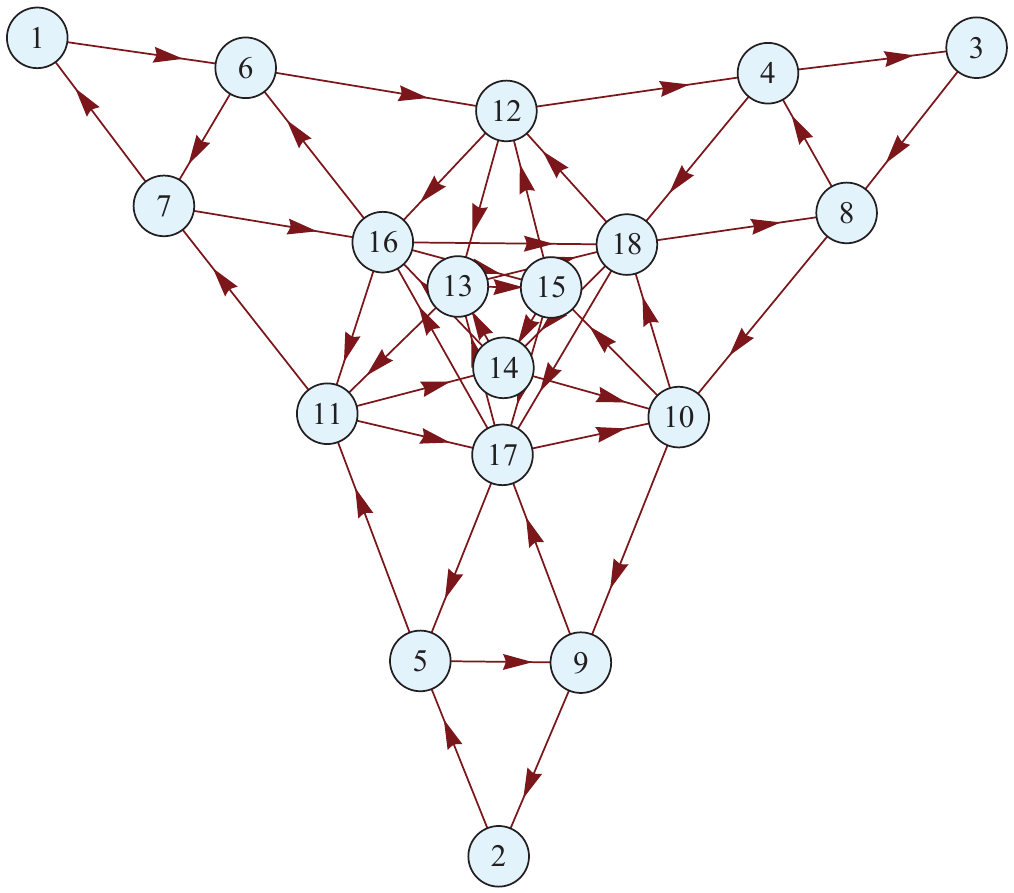}

\includegraphics[width=14.5cm]{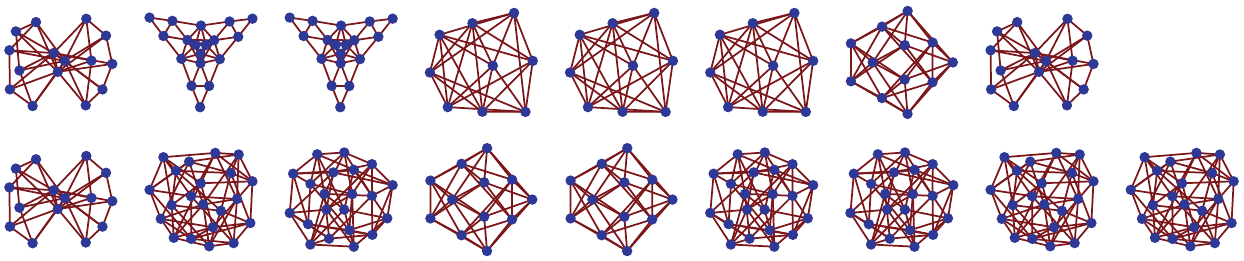}
\caption{Fusion graph $N_6$ of the Drinfeld double of the group $\Sigma_{168}\times\mathbb{Z}_3$.}
\label{fig:Sigma168x3_6}
\end{figure}

Embedding labels: 6, 7 and their conjugates 8, 9.
See the graph of irrep 6 on Fig.~\ref{fig:Sigma168x3_6}.

\subsubsection[Drinfeld double of $\Sigma_{216\times3}$]{Drinfeld double of $\boldsymbol{\Sigma_{216\times3}}$}

The group $\Sigma_{216\times3}$ is called the Hessian group, but this name also refers to $\Sigma_{216}$.

Order of the group: $648$

GAP nomenclature: $\Sigma_{216}\times\mathbb{Z}_3={\rm SmallGroup}(648,532)$

Class number: $\ell=24$

Classical dimensions: {1, 1, 1, 2, 2, 2, 3, 3, 3, 3, 3, 3, 3, 6, 6, 6, 6, 6, 6, 8, 8, 8, 9,9}

Rank: ${r}=486$

$N_c=\{24,21,24,24,27,9,9,12,21,21,27,27,27,27,27,27,12,12, 18,18,18,18,18,18\}$

Quantum dimensions:
$(1_3,2_3,3_7,6_6,8_3,9_2;9_9,18_9,27_3;1_3,2_3,3_7,6_6,8_3,9_2;1_3,2_3,3_7,6_6,8_3,9_2;$

\hspace*{38mm}$24_{27};\!72_{9};\!72_{9};\!54_{12};\!9_9,18_9,27_3;\!9_9,18_9,27_3;\!12_{18},24_{9};\!12_{18},24_{9};\!12_{18},24_{9};$

\hspace*{38mm}$12_{18},24_{9};\!12_{18},24_{9};\!12_{18},24_{9};\!54_{12};\!54_{12};\!36_{18};\!36_{18};\!36_{18};\!36_{18};\!36_{18};\!36_{18})$

$d_{\mathcal B}={{2^2}{3^6}{13^1}{787^1}{1481^1}}$

Embedding labels: (pairwise conjugates) 8, 10, 11 and 9, 13, 12.

\begin{figure}[htp]\centering
\includegraphics[width=4.5cm]{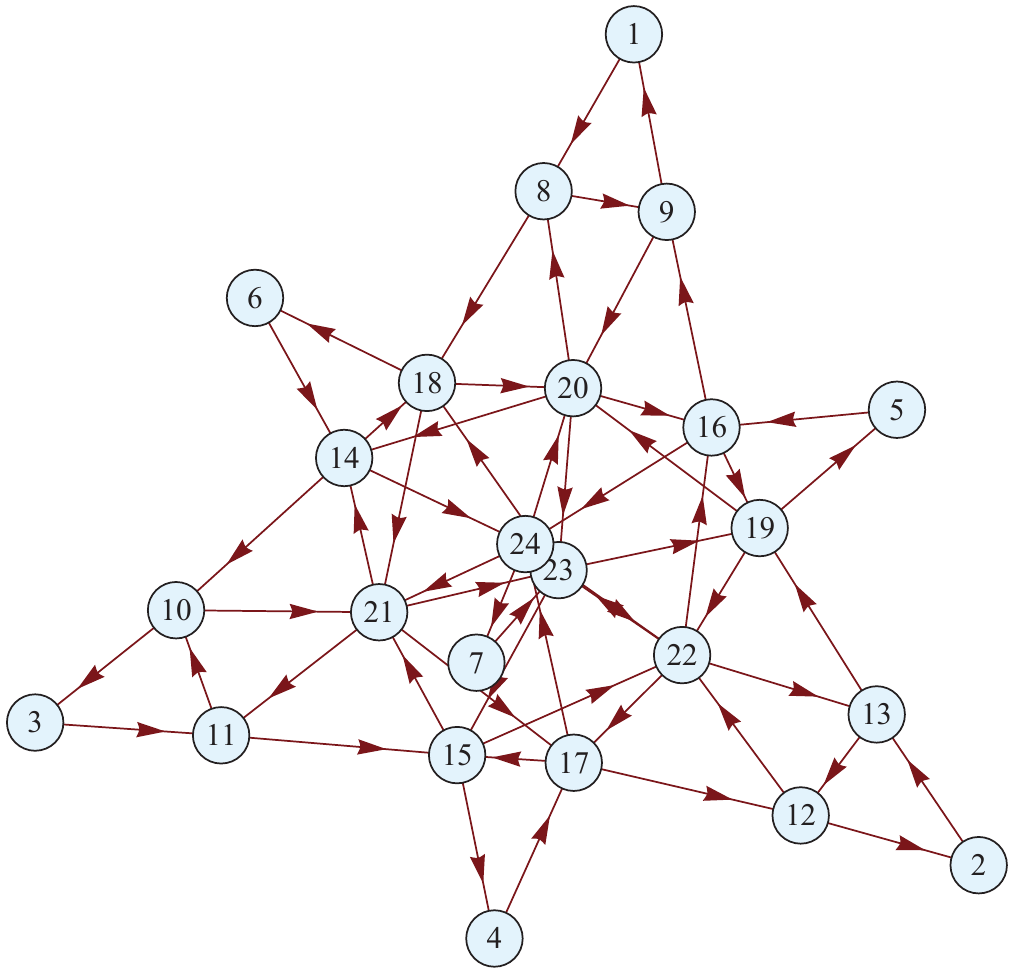}

\includegraphics[width=14.5cm]{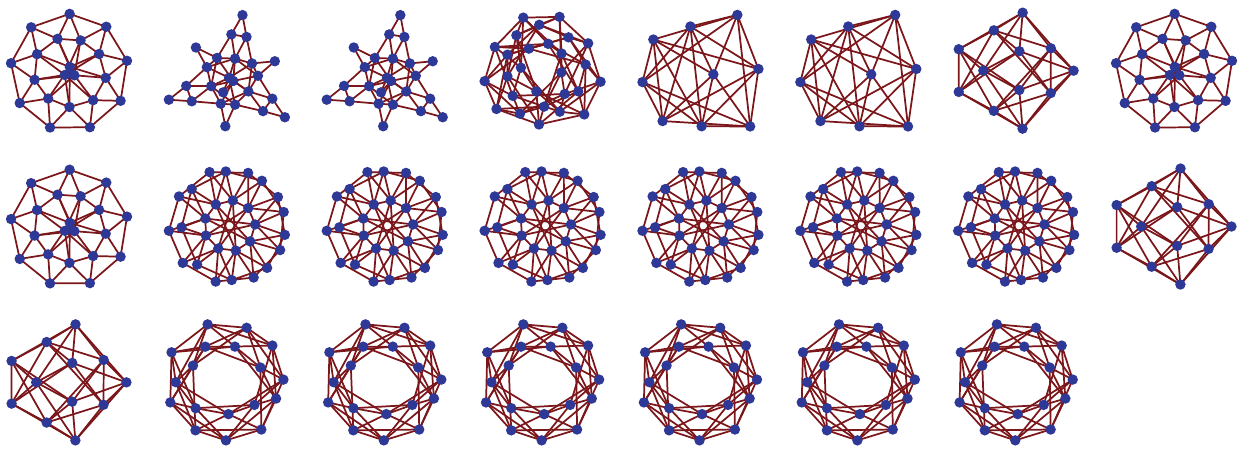}

\caption{Fusion graph $N_8$ of the Drinfeld double of the group  $\Sigma_{216\times3}$.}
\end{figure}

\subsubsection[Drinfeld double of $\Sigma_{360\times3}$]{Drinfeld double of $\boldsymbol{\Sigma_{360\times3}}$}

The group $\Sigma_{360\times3}=\Sigma_{1080}$ (not
always realized as a~subgroup of ${\rm SU}(3)$) is sometimes called the Valentiner group but this name also
refers to $\Sigma_{360}$.

 Order of the group: $1080$

GAP nomenclature: $\Sigma_{360\times3}={\rm SmallGroup}(1080,260)$.
{Alternate names: $3.A_6$.}

Class number: $\ell=17$

Classical dimensions: {1, 3, 3, 3, 3, 5, 5, 6, 6, 8, 8, 9, 9, 9, 10, 15, 15}

Rank: $r=240$

$N_c={17,15,17,17,9,9,12,15,15,15,15,12,12,15,15,15,15}$

Quantum dimensions: $(1,3_4,5_2,6_2,8_2,9_3,10,15_2;45_{12},90_{3};1,3_4,5_2,6_2,8_2,9_3,10,15_2;$

\hspace*{38mm}$1,3_4,5_2,6_2,8_2,9_3,10,15_2;120_9;120_9;90_{12};72_{15};72_{15};45_{12},90_3;$

\hspace*{38mm}$45_{12},90_3;90_{12};90_{12};72_{15};72_{15};72_{15};72_{15})$

$d_{\mathcal B}={{2^1}{3^3}{734267099^1}}$

\begin{figure}[htp]\centering
\includegraphics[width=6.5cm]{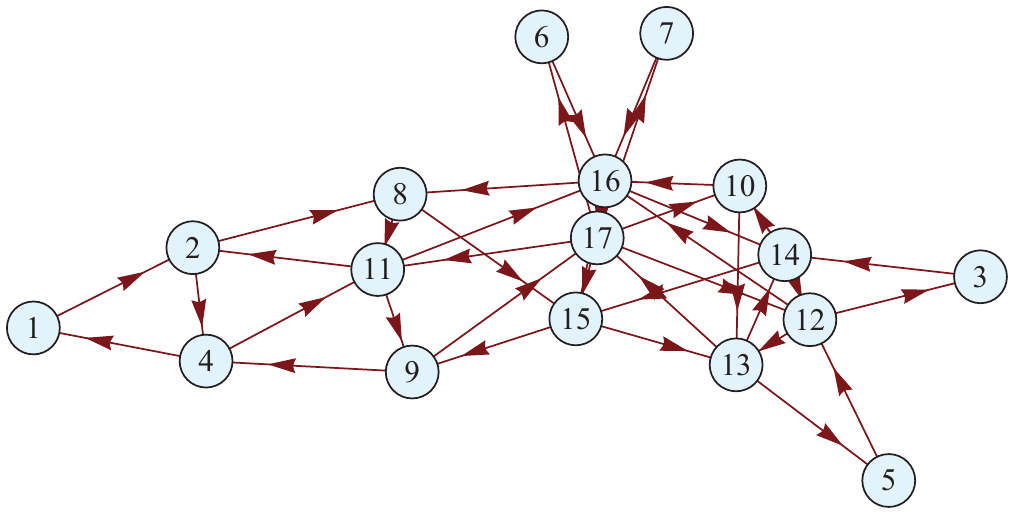}

\includegraphics[width=14.5cm]{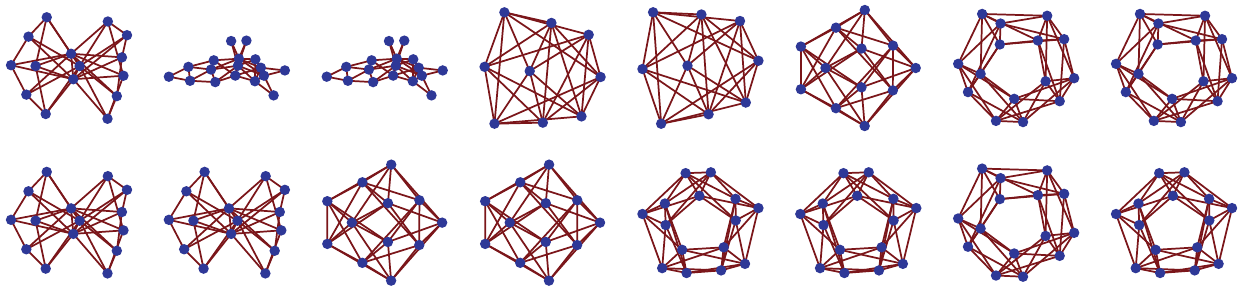}
\caption{Fusion graph $N_2$ of the Drinfeld double of the group $\Sigma_{360\times3}$.}
\label{fig:Sigma1080}
\end{figure}

Embedding labels: 2 and 4.
See the graph of $N_2$ on Fig.~\ref{fig:Sigma1080}.

{\bf Note.} All our results about these subgroups of ${\rm SU}(3)$, the violations of
sumrules~\eqref{sumrule},~\eqref{XeqXC} and~\eqref{sumruleS} and the ``accidental'' vanishings have been
collected in Table~\ref{ssgrSU3}.

\begin{table}[t]\centering
\caption{Data and status of the sumrules~\eqref{sumrule} and~\eqref{sumruleS} for Drinfeld doubles
of some subgroups of ${\rm SU}(3)$.
The meaning of symbols is like in Table~\ref{ssgrSU2}.}
\label{ssgrSU3}

\vspace{1mm}

$
\hskip5mm \noindent\renewcommand{\arraystretch}{1.2}
\begin{array}
{|c||c||c|c|c|c|c|c|}
\hline
{\mbox{ name}} & {\eqref{sumrule} \atop {\rm before\atop doubling}}&r & \# i,\;i \neq {\bar
i},\;\forall\, j\atop \sum\limits_k N_{ij}^k\buildrel{?}\over =\sum\limits_k N_{\bar \imath j}^k & { \#
\;\hbox{complex} \atop \# \;\sum\limits_j S_{ i j} =0 } & { \# \;\hbox{quatern.} \atop \#
\sum\limits_j S_{ i j} =0} & { \# \;\hbox{real} \atop \# \sum\limits_j S_{ i j} =0} & \#
\hbox{units}
\\[4pt]
\hline
\hline
\Delta_{3\times 2^2}=T & \checkmark &14& 2 &8\atop 6\checkmark\;0 A & 0&6\atop 0 &3
\\[0pt]

\hline
\Delta_{6\times 2^2}=O &\checkmark&21& \checkmark &0 & 0 & 21\atop 9\checkmark\;1 A&2
\\[0pt]

\hline
F_{21} &{\times}&25 &2 &24 \atop 6 \checkmark\;0 A & 0 & 1\atop 0 & 3
\\[0pt]

\hline
\hline
\Sigma_{60}=I & \checkmark &22 &\checkmark & 0 &0&22\atop 4\checkmark \;0A & 1
\\[0pt]

\hline
\Sigma_{36\times 3} &\checkmark &168 &\checkmark&156\atop 156\,\checkmark\;14A& 0 &12 \atop 3\checkmark \;1
A & 12
\\[0pt]

\hline
\Sigma_{168} &\checkmark& 32 &2&16\atop 14\,\checkmark\;14 A & 0&16\atop 0 &1
\\[2pt]

\hline
\Sigma_{60}\times \Bbb{Z}_3 &\checkmark &198 & 176 & 176\atop 176\,\checkmark\;0A & 0& 22\atop 4\checkmark
\;4 A & 9
\\[0pt]

\hline
\Sigma_{72\times 3} &\times &210 & 46 &184\atop 162\,\checkmark
\qquad
0 A &8\atop 6\,\checkmark\;0 A&18\atop 6 \checkmark\;0 A &12
\\[0pt]

\hline
\Sigma_{168}\times \Bbb{Z}_3 &\checkmark&288 & 146& 272 \atop 270 \, \checkmark\;14 A &0&16\atop 0 & 9
\\[2pt]

\hline
\Sigma_{216\times 3} &\checkmark& 486 & 472 & 472\atop 472 \,\checkmark\;58A & 4 \atop 4\checkmark\;4 A
\,\checkmark &10 \atop 2\checkmark \;2A&9
\\[1pt]

\hline
\Sigma_{360\times 3} &\times&240 &52 &208\atop 176\,\checkmark\;20A & 0 &32\atop 0 &3
\\[3pt]

\hline
\hline
\end{array}
$
\end{table}

\subsection{Drinfeld doubles of other f\/inite groups (examples)}

We decided, in this paper, to discuss Drinfeld doubles of f\/inite subgroups of ${\rm SU}(2)$ and of ${\rm
SU}(3)$.
But this overgroup plays no role and it could have been quite justif\/ied to organize our results in
a~dif\/ferent manner.
Some f\/inite subgroups of ${\rm SO}(3)$ have been already encountered as subgroups of ${\rm SU}(3)$, see
above the cases of $T\cong\Delta_{3\times2^2}$, $O\cong\Delta_{6\times2^2}$ and $I\cong\Sigma_{60}$.
We have also looked at the eight Mathieu groups, see for instance~\cite{grouppropssite},
$M_9$, $M_{10}$, $M_{11}$, $M_{12}$, $M_{21}$, $M_{22}$, $M_{23}$, $M_{24}$.
Some ($M_9, M_{10}, M_{12}, M_{21}$) satisfy the sumrule~\eqref{sumrule} before doubling, the others
don't, and we were unable to understand the reason behind this.

\vspace{-1mm}

\section{Conclusion and outlook} Along this guided tour of Drinfeld doubles of f\/inite subgroups of
${\rm SU}(2)$ and ${\rm SU}(3)$ we have put the emphasis on the discussion of the ``sum rules'' discovered
in~\cite{RCJBZ:sumrules}: we have found that they are not generally satisf\/ied, just like in the case of
f\/inite groups.
In that respect the modularity of the fusion category of Drinfeld doubles is of no help.
But we also found that, like in simple or af\/f\/ine Lie algebras, the two properties~\eqref{sumrule}
and~\eqref{sumruleS} are equivalent, in the sense that they are simultaneously satisf\/ied for all $i$ and
$j$ or not, a~property that has no equivalent for f\/inite groups, see Proposition~\ref{proposition1}.
We found that certain conditions, like the existence of units, may grant the vanishing of some of these sum
rules (Proposition~\ref{proposition3}).
But a~certain number of observed vanishing sum rules remains with no obvious explanation, whence the name
``accidental'' that we gave them.
The bottom line is that at this stage, we have no precise criterion to decide for which group $G$ and for
which irrep of $D(G)$ the sum rule~\eqref{sumruleS} is satisf\/ied.
Clearly these points should deserve more attention.

The paper contains many f\/igures.
One of the curious properties of the fusion graph of an ``embedding'' representation of $D(G)$~-- the
appropriate generalization of the concept of fundamental faithful representation in this context~-- is that
it contains as many connected components as there are irreps or classes in $G$.

Finally we want to mention another open issue that was not explicitly discussed in the present paper.
It turns out that fusion graphs of doubles of f\/inite subgroups of SU(2) have lots of similarities with
those appearing in the discusion of rational $c=1$ conformal f\/ield theories obtained by orbifolding SU(2)
by these f\/inite subgroups~\cite{CappDAppo,DijkgraafEtAl, Ginsparg, Vafa:discretetorsion}.
The precise relation, however, remains elusive.
We hope to return to this point in the future.


\appendix
\vspace{-1mm}

\section{About representations, faithfulness, and embeddings}\label{appendixA}

As usual, if no further indication is given, by ``representation of a~f\/inite group'' we mean ``linear
representation in a~complex vector space''.
A representation of dimension (or degree) $n$, and kernel $K$, of the (abstract) f\/inite group $G$, is in
particular a~morphism from $G$ to ${\rm GL}(n,\mathbb{C})$ and determines a~concrete realization of the
group~$G\vert K$ (namely the image of this morphism) as a~subgroup of ${\rm GL}(n,\mathbb{C})$.
Conversely, if we choose some concrete f\/inite subgroup of ${\rm GL}(n,\mathbb{C})$, and if this subgroup
is isomorphic to~$G$, the choice determines a~faithful representation of~$G$ (whose kernel $K$ is trivial)
of dimension~$n$.
We shall say that a~faithful $n$-dimensional representation of~$G$ determines an embedding of~$G$ into
${\rm GL}(n,\mathbb{C})$ and that it determines an irreducible embedding if the representation is both
irreducible and faithful.

\looseness=-1
As usual, the table of characters of a~f\/inite group describes the evaluation of irreducible characters
(listed vertically) on conjugacy classes (listed horizontally).
The lack of faithfulness of an (irreducible) representation~$\rho$ of dimension~$n$ can be detected on the
table of characters by the existence, along the line~$\rho$, of more than one entry equal to~$n$; this is
a~necessary and suf\/f\/icient condition, as one may easily convince oneself.
The representations appearing in a~table of characters are irreducible and inequivalent (by def\/inition of
a~table of characters), but a~table may contain several faithful representations of the same dimension.
Such representations therefore def\/ine subgroups of~${\rm U}(n)$ that are isomorphic to the given
f\/inite group, but are not conjugated.

\looseness=-1
As every representation of a~f\/inite group is equivalent to a~unitary representation, we may consider that
a~faithful $n$-dimensional representation of $G$ determines an embedding of $G$ as a~subgroup of ${\rm
U}(n)={\rm U}(n,\mathbb{C})$.
Using the determinant homomorphism $\det$ from ${\rm U}(n)$ to ${\rm U}(1)$, one can consider ${\rm U}(n)$
as a~semi-direct product of ${\rm U}(1)$ and ${\rm SU}(n)$.
From any faithful unitary representation $\rho$ of the f\/inite group $G$,  i.e.,  from any
embedding of~$G$ into~${\rm U}(n)$, one obtains, by composing with the determinant, a~${\rm U}(1)$-valued
representation, which may be trivial or not.
In order to obtain an embedding of~$G$ into ${\rm SU}(n)$ one needs the determinant to be equal to~$1$ on~$\rho(G)$.
Calculating the value of $\det$ for the chosen representation is not immediate from the table of characters
alone, so that one usually needs to make use of an explicit set of representatives for the group generators.

\looseness=-1
An irreducible representation {of~$G$} (in a~complex vector space) can be of real, complex, or quaternionic type.
{If it is faithful and volume preserving, it determines an embedding of~$G$ into~${\rm SU}(n)$, but if it
is furthermore of real type, one can obtain an embedding of $G$ into ${\rm SO}(n)\subset{\rm SU}(n)$, and
if it is of quaternionic type} (the dimension $n$ should be even), it determines an embedding into the
compact group ${\rm Sp}(n/2)={\rm U}(n/2,\mathbb{H}))\subset{\rm SU}(n)$.
Remember that ${\rm Sp}(1)={\rm SU}(2)$.
The type of an irreducible representation is determined by the Frobenius--Schur indicator (not to be
confused with the Schur index), which is~$+1$ for~$\mathbb{R}$, $0$~for~$\mathbb{C}$ and~$-1$ for~$\mathbb{H}$.
In the cases like the Drinfeld doubles in which we have $S$ and $T$ matrices, see~\cite{RCJBZ:sumrules} for
an expression of that Frobenius--Schur indicator.

\section{About primitive and imprimitive f\/inite subgroups\\ of Lie groups}\label{appendixB}

 A f\/inite subgroup $G$ of a~Lie group is
called Lie primitive (we shall just write ``primitive'') if it is not included in a~proper closed Lie
subgroup (i.e.,  not a~discrete subgroup) of that Lie group.
One also says that~$G$ is, or determines, a~primitive inclusion in the chosen Lie group.
If~$G$ is not primitive, it is called imprimitive.
A notion of Lie primitivity also exists for continuous subgroups of Lie groups but we do not need it.
For instance the f\/inite subgroups in ${\rm SO}(3)$ are the cyclic and dihedral groups together with the
three exceptional\footnote{We call ``exceptional'' a~subgroup of a~given Lie group ${\rm SU}(n)$ that is
not member of an inf\/inite series.} polyhedral groups $T=A_4$, $O=S_4$ and $I=A_5$, and they are primitive in ${\rm SO}(3)$ except for the cyclic groups since the latter are subgroups ${\rm U}(1) \cong {\rm SO}(2)\subset{\rm SO}(3)$.
The f\/inite subgroups of ${\rm SU}(2)$ are the cyclic and binary dihedral groups, together with three
exceptional: the binary polyhedral groups $\widehat T$, $\widehat O$ and~$\widehat I$.
The cyclic groups are not primitive in ${\rm SU}(2)$ since they are subgroups of ${\rm U}(1)\subset{\rm
SU}(2)$, but they are not irreducible either as subgroups of ${\rm SU}(2)$ since the restriction to
a~cyclic group of its representation of complex dimension $2$ splits as $1\oplus1$.

\looseness=1
The situation is more subtle for subgroups of ${\rm SU}(3)$.
For instance the existence of one (actually two) irreps of dimension $3$ for the icosahedral group
$\Sigma_{60}$ insures the existence of an irreducible embedding in ${\rm SU}(3)$, but these irreps of real
type ref\/lect the existence of the chain $\Sigma_{60}\subset{\rm SO}(3)\subset{\rm SU}(3)$.
So, although irreducible, the embedding of $\Sigma_{60}$ in ${\rm SU}(3)$ is not primitive.

\section{About Schur covers}\label{appendixC}

Every projective representation of a~f\/inite group $G$, def\/ined as a~representation of $G$ on
a~projective space, can be lifted to an ordinary (i.e., linear) representation of another group
called a~Schur covering group after the seminal work of I.~Schur at the beginning of the $20^{th}$ century
(see~\cite{Curtis-Reiner} and references therein).
Every f\/inite group has a~Schur cover but the latter may be non unique.
As far as the classif\/ication of projective representations of $G$ is concerned, this non uniqueness does
not matter because one can choose any Schur cover: the obtained projective representations of $G$ turn out to be the same.
The case of alternating and symmetric groups is discussed in~\cite{HoffmanHumphrey}, see also~\cite{symcovers:wiki}.

We review a~few basic facts about f\/inite group extensions.
An extension $K$ of $G$ by $N$ is def\/ined by an onto group homomorphism $K\mapsto G$ with kernel $N$.
The latter is a~normal subgroup of $K$ and $G$ is isomorphic with $K|N$.
The extension is called central whenever the kernel lies in the center ${Z}_K$ of $K$, in which case $N$ is abelian.
A stem extension of $G$ is a~central extension such that $N$ is not only contained in the center
${Z}_K$ of $K$ but also in the {commutator} subgroup $K^\prime$ of $K$, so $N\subset {Z}_K\cap K^\prime$.
A Schur covering group (or a~Schur cover) $K$ of $G$ can be abstractly def\/ined as a~stem extension of maximal order.
Usually they are not split\footnote{An extension is called split if $K$ is not only an extension, but also
a~semi-direct product of $N$ and $G$ with respect to some morphism $\phi$ from $G$ to the group ${\rm Aut}(N)$.
One writes $K=N\rtimes_\phi G$, or just $K=N\rtimes G$, but~$\phi$ is needed since the product is def\/ined
as $(n_1,g_1)(n_2,g_2)=(n_1\phi_{g_1}(n_2),g_1g_2)$.
In that case both $N$ and $G$ are subgroups of $K$.
A split extension can be central or not.} extensions ($K$ has no reason to be a~semi-direct product of $N$ and $G$).

Up to isomorphism, the normal subgroup $N$ def\/ining the Schur extension does not depend on the choice of
the chosen Schur covering group $K$ of $G$.
More precisely, $N$ is isomorphic with what is called the Schur multiplier of $G$, def\/ined as
$M(G)=H_2(G,\mathbb{Z})=(H^2(G,\mathbb{C}^\times))^\star$.
Any projective representation of $G$ determines a~$2$-cocycle, and two projectively equivalent
representations dif\/fer multiplicatively by a~coboundary, so that projective representations of $G$ indeed
fall into classes labelled by elements of $M(G)$.
In particular, if the latter is trivial, the projective representations of $G$ are projectively equivalent to the linear ones.

Classifying Schur covers themselves, for a~given $G$, is a~dif\/f\/icult problem, and we shall not discuss it.
Nevertheless it is worth mentioning that if $G$ is perfect (which means that it is equal to its commutator
group $G^\prime$, or equivalently that its abelianization $H_1(G,\mathbb{Z})=G/G^\prime$ is trivial), then
the Schur cover of $G$ is unique, up to isomorphism, and it may be called ``universal cover'', like in the theory of Lie groups.
However, an imperfect group (like the group of the tetrahedron, see below) may have a~unique Schur cover.
Every non-abelian simple group is perfect.

Let us illustrate this discussion with subgroups of the Lie group ${\rm SO}(3)$.
The universal cover of the latter is ${\rm SU}(2)$, and it is a~binary cover, the kernel of the extension being $\mathbb{Z}_2$.
The corresponding lifts, in ${\rm SU}(2)$, of the subgroups of ${\rm SO}(3)$ def\/ine the so-called binary
groups, and the latter are indeed Schur covers of the former, when~$M(G)$ is not trivial, but not all Schur covers of ${\rm SO}(3)$ subgroups lie in ${\rm SU}(2)$.
The Schur multiplier of the exceptional polyhedral groups is $\mathbb{Z}_2$, and this means that they have
two kinds of projective representations: those obtained from the reduction of the ${\rm SO}(3)$-representations (integral spin)
and associated with their usual (linear) representations, but also
the so-called binary representations (they are not linear representations) which can be obtained from the
reduction of the ${\rm SU}(2)$ representations of half-integral spin.
The tetrahedral group $T=A_4$ is not perfect, nevertheless it has a~unique Schur cover\footnote{A central
extension of a~group $G$ by the cyclic group $\mathbb{Z}_k$ is often denoted $k.G$.}, $2.A_4$, that
coincides with the binary tetrahedral group $\widehat T={\rm SL}(2,3)$.
The icosahedral group $I\sim A_5\sim\Sigma_{60}$ is the smallest non abelian simple group, and as such it
is perfect (it is also the smallest non trivial perfect group).
It has therefore only one Schur cover, the binary icosahedral group $\widehat I$, which is itself perfect,
although non-simple, and even ``superperfect'' because $H_1(\widehat I,\mathbb{Z})=H_2(\widehat I,\mathbb{Z})=0$.
However, the octahedral subgroup $O\sim S_4$ of ${\rm SO}(3)$ possesses two non-isomorphic Schur covers,
one is the binary octahedral group $\widehat O$, a~subgroup of ${\rm SU}(2)$, another is the group ${\rm
GL}(2,3)$, which can be realized as a~subgroup of ${\rm U}(2)$, but not of ${\rm SU}(2)$.
Using GAP nomenclature $\widehat O$ can be recognized as SmallGroup(48,28), whereas ${\rm GL}(2,3)$ is SmallGroup(48,29).
From the GAP table of characters only, we see that $\widehat O$ can be embedded into ${\rm SU}(2)$ because
its $2$~dimensional faithful irreducible representations are quaternionic (indeed ${\rm
U}(1,\mathbb{H})\sim{\rm SU}(2)$), whereas the $2$~dimensional faithful irreducible representations of
${\rm GL}(2,3)$ are complex.

\section[The $S$ matrix of the Drinfeld double of the Klein group $\Sigma_{168}$]{The $\boldsymbol{S}$ matrix of the Drinfeld double of the Klein group $\boldsymbol{\Sigma_{168}}$}\label{appendixD}

The blocks of the $32\times32$ symmetric and unitary matrix $168\times S(I,J)$, for $I=1,\ldots,6$, $I\leq
J\leq6$ are given below.
{$(I,J)$} is a~rectangular block, of dimension $u\times v$, with $u$, $v$ taken from the list
$Nc=(6,5,3,4,7,7)$ giving the orders of the centralizers for the conjugacy classes of $\Sigma_{168}$.  We set\footnote{The entries of $S$ lie in the cyclotomic f\/ield
$\mathbb{Q}(\exp(2i\pi/m))$ where $m$ is the exponent of the group~\cite{CGR:modulardata}.
Here $m=84$.
There are many ways to express the same matrix elements.}  $\zeta=-\exp(i\pi/7)$ and
$\xi_{x_0,x_1,x_2,x_3,x_4,x_5}={x_0}+{x_1}\,\xi+{x_2}\,\xi^2+{x_3} \xi^3+{x_4} \xi^4+{x_5} \xi^5$.
${\alpha_1}=\xi_{0,0,0,0,1,2}$, ${\alpha_2}=\xi_{0,0,1,1,1,1}$, ${\alpha_3}=\xi_{0,0,2,1,0,0}$, ${\alpha_4}
=\xi_{0,1,0,2,0,0}$, ${\alpha_5}=\xi_{0,1,1,0,1,0}$, ${\alpha_6}=\xi_{0,2,0,0,0,1}$,
${\alpha_7}=\xi_{1,0,0,1,1,0}$, ${\alpha_8}=\xi_{1,0,1,0,0,1}$, ${\alpha_9}=\xi_{1,1,1,0,1,0}$, ${\alpha_{10}}=\xi_{1,1,1,1,-1,1}$, ${\alpha_{11}}=\xi_{2,2,1,2,2,2}$.
Then,

 $I=1$,
\begin{gather*}
\left(
\begin{matrix}
1&3&3&6&7&8
\\
3&9&9&18&21&24
\\
3&9&9&18&21&24
\\
6&18&18&36&42&48
\\
7&21&21&42&49&56
\\
8&24&24&48&56&64
\\
\end{matrix}
\right),\quad 21\left(
\begin{matrix} 1&1&1&1&2
\\
-1&-1&-1&-1&-2
\\
-1&-1&-1&-1&-2
\\
2&2&2&2&4
\\
-1&-1&-1&-1&-2
\\
0&0&0&0&0
\\
\end{matrix}
\right),\quad 56\left(
\begin{matrix}1&1&1
\\
0&0&0
\\
0&0&0
\\
0&0&0
\\
1&1&1
\\
-1&-1&-1
\\
\end{matrix}
\right),\\
42\left(
\begin{matrix}1&1&1&1
\\
1&1&1&1
\\
1&1&1&1
\\
0&0&0&0
\\
-1&-1&-1&-1
\\
0&0&0&0
\\
\end{matrix}
\right),
\quad
24\left(
\begin{matrix}1&1&1&1&1&1&1
\\
\alpha5&\alpha5&\alpha5&\alpha5&\alpha5&\alpha5
&\alpha5
\\
-\alpha9&-\alpha9&-\alpha9&-\alpha9&-\alpha9&-\alpha9
&-\alpha9
\\
-1&-1&-1&-1&-1&-1&-1
\\
0&0&0&0&0&0&0
\\
1&1&1&1&1&1&1
\\
\end{matrix}
\right),\\
24\left(
\begin{matrix}1&1&1&1&1&1&1
\\
-\alpha9&-\alpha9&-\alpha9&-\alpha9&-\alpha9&-\alpha9
&-\alpha9
\\
\alpha5&\alpha5&\alpha5&\alpha5&\alpha5&\alpha5
&\alpha5
\\
-1&-1&-1&-1&-1&-1&-1
\\
0&0&0&0&0&0&0
\\
1&1&1&1&1&1&1
\\
\end{matrix}
\right),
\end{gather*}

 $I=2$,
\begin{gather*}
21\left(
\begin{matrix}5&1&-3&1&-2
\\
1&5&1&-3&-2
\\
-3&1&5&1&-2
\\
1&-3&1&5&-2
\\
-2&-2&-2&-2&4
\\
\end{matrix}
\right),\quad \left(
\begin{matrix}0&0&0
\\
0&0&0
\\
0&0&0
\\
0&0&0
\\
0&0&0
\\
\end{matrix}
\right),\quad 21\left(
\begin{matrix}2&-2&2&-2
\\
-2&2&-2&2
\\
2&-2&2&-2
\\
-2&2&-2&2
\\
0&0&0&0
\\
\end{matrix}
\right),
\\
\left(
\begin{matrix}0&0&0&0&0&0&0
\\
0&0&0&0&0&0&0
\\
0&0&0&0&0&0&0
\\
0&0&0&0&0&0&0
\\
0&0&0&0&0&0&0
\\
\end{matrix}
\right),\quad \left(
\begin{matrix}0&0&0&0&0&0&0
\\
0&0&0&0&0&0&0
\\
0&0&0&0&0&0&0
\\
0&0&0&0&0&0&0
\\
0&0&0&0&0&0&0
\\
\end{matrix}
\right),
\end{gather*}

 $I=3$,
\begin{gather*}
56\left(
\begin{matrix}2&-1&-1
\\
-1&-1&2
\\
-1&2&-1
\\
\end{matrix}
\right),\quad\left(
\begin{matrix}0&0&0&0
\\
0&0&0&0
\\
0&0&0&0
\\
\end{matrix}
\right),\quad \left(
\begin{matrix}0&0&0&0&0&0&0
\\
0&0&0&0&0&0&0
\\
0&0&0&0&0&0&0
\\
\end{matrix}
\right),\\
\left(
\begin{matrix}0&0&0&0&0&0&0
\\
0&0&0&0&0&0&0
\\
0&0&0&0&0&0&0
\\
\end{matrix}
\right),
\end{gather*}

 $I=4$,
\begin{gather*}
84\left(
\begin{matrix}1&0&-1&0
\\
0&-1&0&1
\\
-1&0&1&0
\\
0&1&0&-1
\\
\end{matrix}
\right),\quad\left(
\begin{matrix}0&0&0&0&0&0&0
\\
0&0&0&0&0&0&0
\\
0&0&0&0&0&0&0
\\
0&0&0&0&0&0&0
\\
\end{matrix}
\right),\quad \left(
\begin{matrix}0&0&0&0&0&0&0
\\
0&0&0&0&0&0&0
\\
0&0&0&0&0&0&0
\\
0&0&0&0&0&0&0
\\
\end{matrix}
\right),
\end{gather*}

$I=5$,
\begin{gather*}
24\left(
\begin{matrix}3&-{\alpha_9}&-{\alpha_9}&{\alpha_5}&-{\alpha_9}&{\alpha_5}&{\alpha_5}
\\
-{\alpha_9}&{\alpha_3}&{\alpha_6}&-{\alpha_2}&-{\alpha_{10}}&{\alpha_8}&{\alpha_7}
\\
-{\alpha_9}&{\alpha_6}&-{\alpha_{10}}&{\alpha_7}&{\alpha_3}&-{\alpha_2}&{\alpha_8}
\\
{\alpha_5}&-{\alpha_2}&{\alpha_7}&-{\alpha_{11}}&{\alpha_8}&{\alpha_1}&{\alpha_4}
\\
-{\alpha_9}&-{\alpha_{10}}&{\alpha_3}&{\alpha_8}&{\alpha_6}&{\alpha_7}&-{\alpha_2}
\\
{\alpha_5}&{\alpha_8}&-{\alpha_2}&{\alpha_1}&{\alpha_7}&{\alpha_4}&-{\alpha_{11}}
\\
{\alpha_5}&{\alpha_7}&{\alpha_8}&{\alpha_4}&-{\alpha_2}&-{\alpha_{11}}&{\alpha_1}
\\
\end{matrix}
\right),\\
24\left(
\begin{matrix}3&{\alpha_5}&{\alpha_5}&-{\alpha_9}&{\alpha_5}&-{\alpha_9}&-{\alpha_9}
\\
{\alpha_5}&{\alpha_1}&-{\alpha_{11}}&-{\alpha_2}&{\alpha_4}&{\alpha_8}&{\alpha_7}
\\
{\alpha_5}&-{\alpha_{11}}&{\alpha_4}&{\alpha_7}&{\alpha_1}&-{\alpha_2}&{\alpha_8}
\\
-{\alpha_9}&-{\alpha_2}&{\alpha_7}&{\alpha_6}&{\alpha_8}&{\alpha_3}&-{\alpha_{10}}
\\
{\alpha_5}&{\alpha_4}&{\alpha_1}&{\alpha_8}&-{\alpha_{11}}&{\alpha_7}&-{\alpha_2}
\\
-{\alpha_9}&{\alpha_8}&-{\alpha_2}&{\alpha_3}&{\alpha_7}&-{\alpha_{10}}&{\alpha_6}
\\
-{\alpha_9}&{\alpha_7}&{\alpha_8}&-{\alpha_{10}}&-{\alpha_2}&{\alpha_6}&{\alpha_3}
\\
\end{matrix}
\right),
\end{gather*}

 $I=6$,
\begin{gather*}
24\left(
\begin{matrix}3&-{\alpha_9}&-{\alpha_9}&{\alpha_5}&-{\alpha_9}&{\alpha_5}&{\alpha_5}
\\
-{\alpha_9}&{\alpha_3}&{\alpha_6}&-{\alpha_2}&-{\alpha_{10}}&{\alpha_8}&{\alpha_7}
\\
-{\alpha_9}&{\alpha_6}&-{\alpha_{10}}&{\alpha_7}&{\alpha_3}&-{\alpha_2}&{\alpha_8}
\\
{\alpha_5}&-{\alpha_2}&{\alpha_7}&-{\alpha_{11}}&{\alpha_8}&{\alpha_1}&{\alpha_4}
\\
-{\alpha_9}&-{\alpha_{10}}&{\alpha_3}&{\alpha_8}&{\alpha_6}&{\alpha_7}&-{\alpha_2}
\\
{\alpha_5}&{\alpha_8}&-{\alpha_2}&{\alpha_1}&{\alpha_7}&{\alpha_4}&-{\alpha_{11}}
\\
{\alpha_5}&{\alpha_7}&{\alpha_8}&{\alpha_4}&-{\alpha_2}&-{\alpha_{11}}&{\alpha_1}
\\
\end{matrix}
\right).
\end{gather*}

\hspace*{82mm}\href{http://math.ucr.edu/home/baez/klein.html}{Klein's quartic} \quad
\raisebox{-16mm}[0pt][0pt]{\includegraphics[scale=0.40]{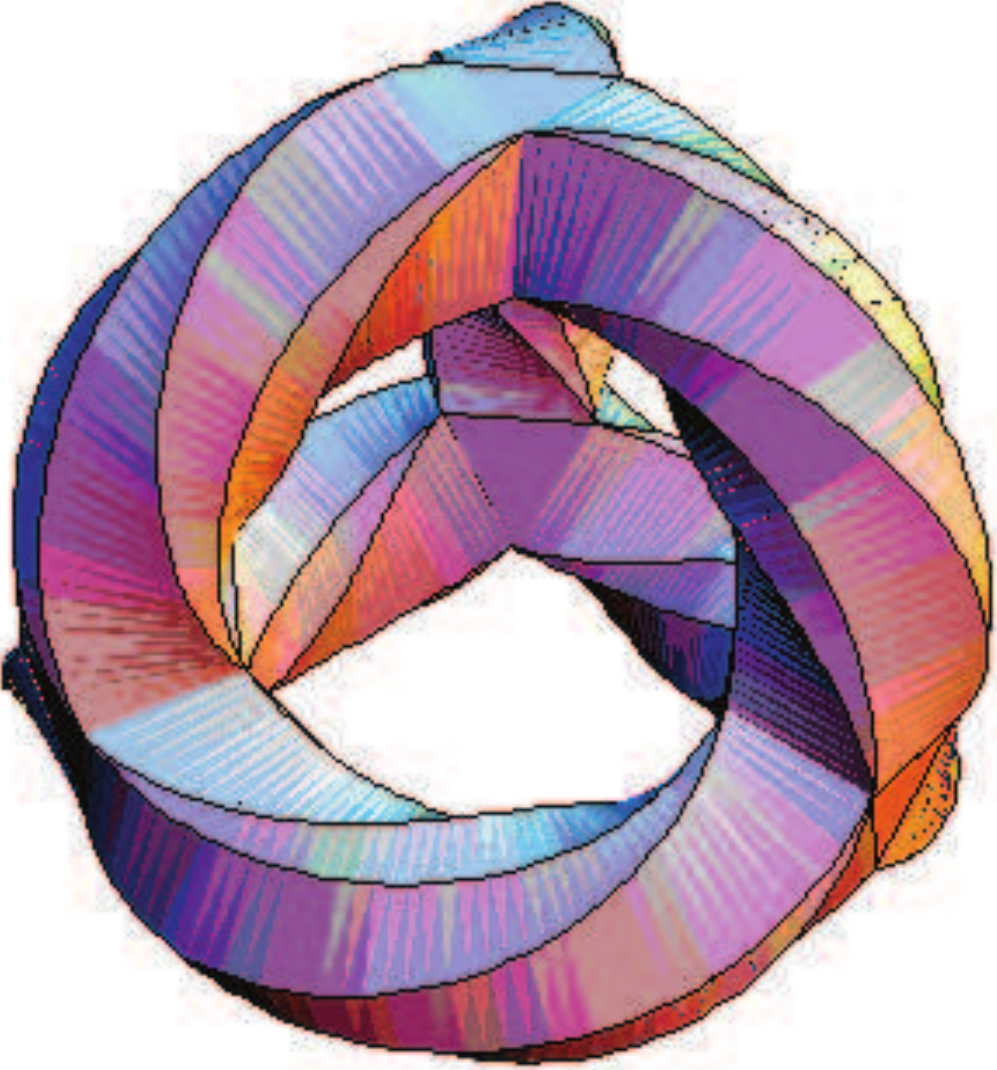}}


\pdfbookmark[1]{References}{ref}
\LastPageEnding

\end{document}